 \documentclass[aps, twocolumn, showpacs, amsmath]{revtex4}
\usepackage{dcolumn}
\usepackage{bm}
\usepackage{graphicx}
\usepackage{color}
\usepackage{subfigure}
\usepackage{hyperref}
\usepackage{latexsym}
\usepackage{amsthm}
\usepackage{amssymb}
\usepackage{comment}

\DeclareGraphicsExtensions{.jpg,.pdf, .mps, .png, .eps, .ps, .EPS,.gif}

\begin{document}
\def\be{\begin{equation}}
\def\ee{\end{equation}}

\def\bc{\begin{center}}
\def\ec{\end{center}}
\def\bea{\begin{eqnarray}}
\def\eea{\end{eqnarray}}
\newcommand{\avg}[1]{\langle{#1}\rangle}
\newcommand{\Avg}[1]{\left\langle{#1}\right\rangle}

\def\ie{\textit{i.e.}}
\def\etal{\textit{et al.}}
\def\m{\vec{m}}
\def\G{\mathcal{G}}

\newcommand{\filippo}[1]{{\bf\color{blue}#1}}
\newcommand{\gin}[1]{{\bf\color{blue}#1}}

\newcommand{\change }[1]{#1}

\title{Redundant  interdependencies boost the  robustness of  {multiplex} networks}

\author{ Filippo Radicchi }
\affiliation{Center for Complex Networks and Systems Research, School
  of Informatics and Computing, Indiana University, Bloomington, IN
  47408, USA}
\author{Ginestra Bianconi}
\affiliation{School of Mathematical Sciences, Queen Mary University of London, London, E1 4NS, United Kingdom}

\begin{abstract}
 {In the analysis of the robustness of multiplex networks, it is
  commonly assumed that}
a node is functioning only if  {its interdependent nodes} are  simultaneously functioning. According to this model,
a  multiplex network becomes 
more and more fragile as the number 
of layers increases. In this respect, the addition of a new 
layer of interdependent
nodes to a preexisting multiplex network will never improve its
robustness. Whereas such a model seems appropriate 
to understand the effect of interdependencies
in the simplest scenario of a network composed of only two layers,
it may seem not  {suitable to characterize the robustness} of real systems formed by 
multiple network layers. 
It seems in fact  {unrealistic} that a real system, 
evolved, through the development
of multiple layers of interactions,  towards  a fragile structure.
In this paper, we introduce a  model of percolation 
where the condition that makes
a node functional is that the node is functioning
in at least two of the layers of the network.
The model reduces to the 
 {commonly adopted}
percolation model
for multiplex networks when the number of layers equals two.
For larger number of layers, however, the model describes
a scenario where the addition of new layers
boosts the robustness of the system by
creating redundant interdependencies among layers.
We prove this fact thanks to the development of a message-passing  theory
able to characterize the model in both synthetic and real-world
multiplex graphs.
\end{abstract}


\maketitle

\section{Introduction}

Multilayer networks~\cite{PhysReports,
  Kivela, Goh_review, Havlin1, gao2012} are emerging as a powerful  
paradigm for describing complex systems 
characterized by the
coexistence of different types of interactions
 {or coupling among different types of networks.}
Multilayer networks represent an appropriate
descriptive model for real networked systems
in disparate contexts, such as  social \cite{Thurner,Weighted},
 technological \cite{Boccaletti,Arenas_t,Vito} 
and biological systems \cite{Menche,Bullmore2009,Makse} {, just
  to mention a few of them.}
For example,
global infrastructures are formed 
by several interdependent networks, such as power grids, water 
supply networks, and communication systems~\cite{Havlin1}.
Cell function and/or malfunction (yielding diseases) cannot be understood if the information 
on the different nature of the interactions forming the interactome (protein-protein interactions, signaling, regulation) are not integrated in a general multilayer scenario~\cite{Menche}.
Similarly, the complexity of the brain is encoded in 
the different
nature of the interactions existing at 
the functional and the structural 
levels~\cite{Bullmore2009, Makse}.

 {General  multilayer networks as well as multiplex networks~\cite{PhysReports,  Kivela,
    Goh_review,Math,Mucha,PRE} are composed by
  nodes belonging to different layers  and links connecting nodes
  within and across layers.  In general multilayer networks  there are
  no restriction in the way the  links across different layers can be
  placed. In multiplex networks, however, the nodes of each pair of
  layers are mapped one to one and the links across different layers
  can only be present among corresponding nodes.  Therefore  multiplex
  networks are  a specific case of multilayer networks with a  well
  defined structure. 
   In multilayer networks, and multiplex networks as well, nodes
  belonging to  different layers are often interdependent on each other, in
  the sense that a failure of one node might cause the failure of
  a node in another layer. For example in global a infrastructure
  network, a power plant might be interdependent on a node of the
  communication system that is controlling its function \cite{Havlin1}.

 Percolation models are
generally used as proxies 
to quantify the robustness of networked 
systems under local failures, by monitoring how the connectedness 
at the macroscopic level changes as a function of the amount of
microscopic  damages of the individual elements of the system. Although different
percolation
models can be suitably defined and studied on multilayer
networks (e.g. k-core percolation~\cite{azimi}, weak
percolation~\cite{baxter}, and bond percolation ~\cite{Hackett16}),
here we focus our attention on the
case of the so-called site-percolation model, where the units
that can be potentially damaged are the nodes of the network
and the order parameter of interest is the macroscopic connectedness
of the system. 
When interdependencies are taken into account, 
the resulting percolation theory \cite{Havlin1,Havlin2,HavlinEPL,Baxter2012,Kabashima,Goh,Son,
  Cellai2016,BD1,BD2,Radicchi2014}
provides a general framework to characterize the dramatic avalanches
of failure that can affect multilayer networks.
A large volume of publications has pointed out that, according to this
  model, a multilayer network is much more fragile than 
each of the various network layers taken in 
isolation~\cite{Havlin1,Havlin2,HavlinEPL,Son,Baxter2012,Kabashima,Goh}.} 
In particular, the overall fragility of the system 
increases as the number of layers increases~\cite{Son, Cellai2016,BD1,BD2,Radicchi2014}. 
Such a feature has an intuitive explanation. 
 {In this percolation model, 
 a node is damaged if 
at least one of its interdependent nodes 
is damaged. As the number of layers increases, the probability 
of individual failures grows thus making the
system more fragile. This scenario leads, however, 
to the conundrum: if
the fragility of a system is increased 
by the number of layers of
interactions, why are there so many real systems that 
display multiple
layers of interactions? Further, 
the addition of new layers of interactions in a pre-existing multilayer network
has generally a cost, so it doesn't seem reasonable 
to spend resources just to make 
the system less robust. 
 {For instance, according to a recent study focusing on diffusion dynamics on
multimodal transportation networks~\cite{Arenas_t}, the presence of
multiple interconnected modes of transportation makes the system
more navigable and more robust than its individual layers considered
in isolation. However, the aforementioned conundrum has not been fully
addressed in terms of purely topological properties.}
The purpose of the current paper is to provide 
a potential explanation by introducing a suitable model for
percolation in multiplex networks composed of multiple interacting layers. 
In the model, we will assume that  a node is damaged only
if all its interdependent nodes are simultaneously damaged. 
The model is perfectly equivalent to the one  {currently in use} when the number
of layer equals two. Additional layers, however, provide the system
with redundant interdependencies, generating backup mechanisms
against the failure of the system, and thus making it more robust.}

 \change{Percolation in absence of interdependencies has been studied  
in multilayer networks finding 
a  ``complementary'' and synergistic role of different
layers~\cite{Leicht09, Lee12, Makse,Hackett16, shao2011}. }
Here, these effects are observed despite of the presence of interdependencies. We provide  a comprehensive 
study of percolation in presence of redundant interdependencies thanks to the development of an exhaustive
message-passing theory~\cite{Kabashima,Lenka,crit,Mezard,Weigt} (also known as the  cavity method). 
We build on recent  advances obtained in the ``standard''
percolation theory  for multilayer networks~\cite{Baxter2016,Cellai2016, Goh_comment,
  Cellai2013,Radicchi} to propose a theory that is valid for
arbitrary systems (thus also including overlap among layers~\cite{PRE, Weighted} ), as long as the
network structure is locally tree-like.
This limitation is common to all message-passing approaches for
studying critical phenomena on networks.  Corrections have been
recently proposed \cite{Radicchi_beyond} on single networks to improve
the performance of message-passing theory and similar approximations
valid for loopy multilayer networks might be envisaged in the future.
 {We remark that our model represents a starting
  point to address an obvious-yet-neglected feature that makes
  percolation more realistic as a model to study the robustness of real multilayer
  networks. Eventual modifications and/or the addition of further
  ingredients  to the model presented here may be still necessary
  to deal with specific scenarios to make the model even more realistic.}

\section{Redundant percolation model on  multiplex networks}

 {
  A multilayer network structure is not
  equivalent to a large single network. As a network is ultimately a
  way to encode information about a complex system, distinguishing
  between different types of links and nodes may significantly alter the
  characteristics of the structure 
and the dynamical behavior of the system.
This fact is particularly evident when we associate a different role to links within each layer (intralinks) and links across different layers (interlinks).
For example when describing the diffusion in a multilayer
networks~\cite{Diffusion}, we might reasonably assume that the
diffusion constant along intralinks 
is different from the diffusion constant among interlinks 
and this changes significantly the typical timescale of the dynamics.
Similar considerations are also valid for spreading processes in
multilayer networks~\cite{Diffusion2}.
In the context of percolation theory, attributing a different role to interlinks and intralinks can yield a scenario significantly different from percolation in single layers.
Specifically, if interlinks describe interdependencies between the nodes, the percolation transition becomes discontinuous and hybrid \cite{Havlin1,Havlin2, Baxter2012}, and close to the percolation transition the multilayer network is affected by large avalanches of cascading failures. In this case, we will refer to the multilayer network as a set of interdependent networks. Therefore ``interdependent networks'' is a term that refers specifically to  the response of the system to the damage of the nodes more than to the actual structure of the multilayer network.

Depending on type and number of interlayer
  connections present in the system, different classes of multilayer networks \cite{PhysReports,Kivela,Goh_review,Math} can
  be considered. Here, we deal with one of the simplest classes
  characterized by the fact that every node is connected to one and 
   only one node in every of the
  other layers. These systems are generally named as ``multiplex
  networks''~\cite{PhysReports,Kivela, Goh_review,Mucha, Math,PRE}. Very often the linked nodes across different layers (also called replica nodes) actually describe different realization of the same node. For example in the London transportation network, replica nodes could represent Oxford Circus tube station and Oxford Circus bus station.  
 \change{ However, multiplex networks can be also used  to indicate the scenario where nodes  belonging to the various layers represent physically distinct units as long as the nodes of the different layers are mapped one-to-one and the links  cross layers are placed among all the corresponding (replica) nodes and nowhere else.
For instance multiplex networks can be used to model 
 interdependent infrastructures when  the interdependencies  occur
 exclusively between replica nodes. Systems of this type are {\em
   interdependent multiplex networks} ~\cite{Havlin1,Baxter2012,Goh, gao2012b} also named  {\em one-to-one interdependent networks}.}

We consider a  {multiplex} network  
composed of  $M$ layers 
$G_{\alpha}$ with $\alpha=1, 2, \ldots, M$. 
We indicate the set of all layers as $\vec{G} = (G_1, G_2, \ldots , G_M)$.
Every layer contains $N$ nodes.
Exactly one node with the same label
appears in every individual layer.
Nodes in the various layers sharing
a common label are called {\it replica nodes}, 
and they are considered as
interdependent on
each other. Nodes in the network are identified  by a pair of
labels $(i,\alpha)$,
with $i=1, 2, \ldots, N$ and $\alpha=1, 2, \ldots, M$, the first one
indicating the index of the node, and the second one standing for the
index of the layer. For every node label $i$, 
the set of replica nodes
is given by the $M$ nodes corresponding to pairs
of labels $(i, \alpha)$ with
with $\alpha=1, 2, \ldots, M$ (see Figure~\ref{fig:multiplex2}). 
When  at least two replica nodes $(i,\alpha)$ and $(i,\alpha')$ are
connected 
to  two corresponding replica nodes $(j,\alpha)$ and $(j,\alpha')$ we 
say that the multiplex network displays link overlap.

\begin{figure}[htb]
\begin{center}
  \includegraphics[width=0.7\columnwidth]{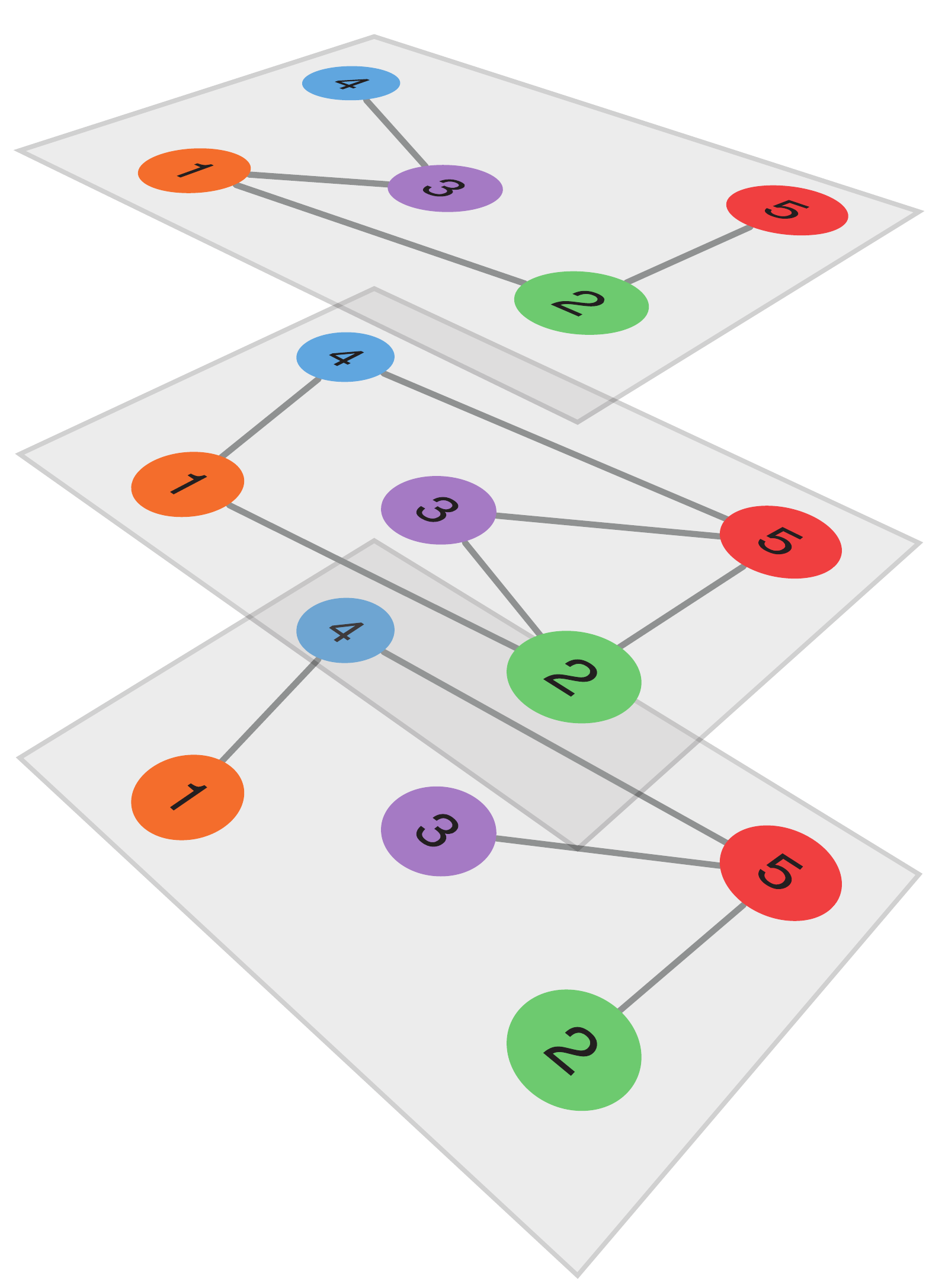}
	\caption{A  {multiplex} network with $M=3$ layers,
          and $N=5$ nodes is shown. Every node $i$ has $M=3$
          interdependent replica nodes $(i,\alpha)$ with
          $\alpha=1, 2, 3$. In this figure, triplets of replica nodes are
also identified by their color.}
	\label{fig:multiplex2}
\end{center}
\end{figure}

Given a  { multiplex } network as described above, we consider
a percolation model where some of the nodes 
are initially damaged. 
\change{We assume that interlinks 
represent interdependencies among replica nodes, but
we consider the case in which such interdependencies 
are {\em redundant}, i.e., every
node can be active only
if at least one its interdependent nodes is also 
active.} We refer to this  model as 
``Redundant percolation model.''
As an order
parameter for the model, we define the so-called
Redundant Mutually Connected Giant Component (RMCGC). 
The nodes that belong to the RMCGC 
can be found by following the algorithm:
\begin{itemize}

\item[(i)]   
The giant component of each layer $\alpha$ is determined, 
evaluating the effect of the damaged nodes in 
each single layer;

\item[(ii)] 
{\em Every replica node that has no other replica node in the giant
  component of its proper layer is removed
from the network and considered as damaged};

\item[(iii)]  
If no new damaged nodes are found at step (ii),
then the algorithm stops, otherwise it proceeds, starting again 
from step~(i).

\end{itemize}
The set of replica nodes that are not damaged when the algorithm stops 
belongs to the RMCGC.

The main difference with the 
percolation model  {introduced in Ref.}~\cite{Havlin1}, and
the consequent definition of Mutually Connected Giant
Component (MCGC),
is that  step (ii) must be substituted with 
``Every replica node that has at least a single replica 
node not in the giant component of its proper layer 
is removed
from the network and considered as damaged, i.e.,  
if a replica node is damaged all its 
interdependent replica nodes are damaged'' ~\cite{Havlin1, Son, Baxter2012,Goh,Kabashima,BD1,BD2,Cellai2016}.
In particular, the RMCGC and the MCGC are 
the same for $M=2$ layers, but they differ
as long as the number of layers $M>2$. 
In the latter case, the RMCGC naturally introduces 
the notion of redundancy  {or complementarity} among 
interdependent nodes. 

\change{Note that the choice of considering just two operating layers, i.e. assumung that  a replica node can remain functional as long as there is at   least another interdependent replica node that is also functional, is a simplification. In a real scenario redundant interdependencies can include more than two operating layers or even a different number of operating layers for each node. }

\begin{figure}[!htb]
\begin{center}
  \includegraphics[width=\columnwidth]{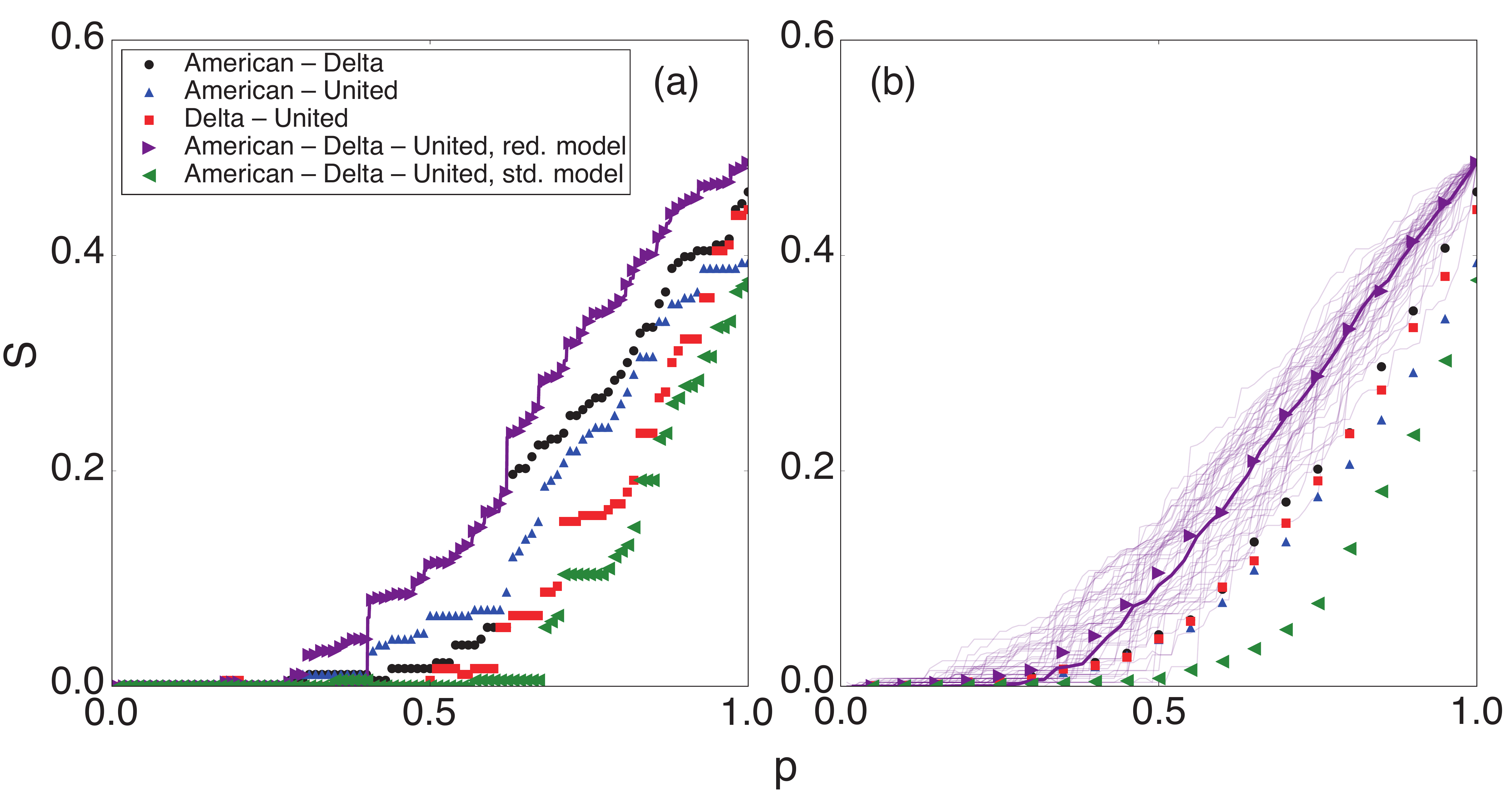}
        \caption{Percolation transition in the US air transportation
          network~\cite{Radicchi}. We consider only US domestic flights operated in
          January 2014 by the three major carriers in the US (American
          Airlines, Delta, and United), and construct a  {multiplex} network where
          airports are nodes, and connections on the layers are
          determined by the existence of at least one flight operated
          by a given carrier between
          the two locations. The number of nodes in the network is
          $N = 183$. Please note that some of the nodes appear as
          connected only in one layer, therefore the  {relative} size of the largest cluster
          is always smaller than one. (a) We consider a single
          realization of the random damage by assigning to every
          replica node $(i, \alpha)$ a random number extracted
          uniformly in the interval $[0,1]$. Replica nodes are
          considered as damaged if their associated random number is
          smaller than $p$. Note that the same exact configuration of random
          damage is considered in all cases. In the percolation
          diagram, we consider the size of the MCGC and 
         RMCGC as a function of $p$. Large symbols are results of numerical simulations,
          whereas the tick line is obtained from the solution of our
          mathematical framework [system of
          Eqs.~(\ref{m1}),~(\ref{v}),~(\ref{s1}), and~(\ref{S})]. 
          In particular, we consider the
          three possible  {multiplex} networks composed of only two
          layers: American -- Delta (black circles), American --
          United (blue triangles), and Delta -- United (red
          squares). Diagrams obtained by
          considering the system as composed of all three
          layers are represented as green triangles for the
          model of Ref.~\cite{Havlin1}, and as purple triangles for the
          redundant model. 
          (b) Same as in panel a, but for average values over $100$
          independent realizations of the random damage. Purple thin
          lines stand for the $100$ independent realizations of random
          damage considered in the case of the redundant model.
        As it is apparent from the  figure, the average
        doesn't well capture the behaviour of single instances of
        disorder and fluctuations are rather large for a wide range
        of possible values of $p$.}
        \label{fig:fig2}
\end{center}
\end{figure}

 {As a proof of concept to demonstrate the difference between 
the notions of MCGC and RMCGC,
in Figure~\ref{fig:fig2} we present
results of numerical simulations for the two percolation models
applied to the air transportation network within the
US~\cite{Radicchi}.  We remark that this analysis represents only
a starting point to illustrate the key impact of the setting 
of redundant interdependency. 
Airports are the nodes in network. Two airports are connected if at
least a flight is connecting them.
Layers correspond to 
flights operated
by the three major carriers in the US: American Airlines, Delta, and
United. 
In this system, having a connected component
that is shared by at least two carriers
 could be important for several reason. For example,
competition on similar itineraries may favor a market for plane
tickets that is more fair than the one that would be
present in case of monopoly by a single
carrier. If only two layers are considered, 
MCGC and RMCGC coincide and both metrics indicate the relative size of the
system where no monopoly is present. The addition of a third layer
in the system should have beneficial effects for the 
system by increasing the size of the system where monopoly is absent.
This scenario is described by the redundant percolation
model. Beneficial effects 
of the complementarity among routes offered by the various carriers are not only visible when the
system is fully functional (i.e., parameter $p=1$), but also when a
relatively large fraction of airports are considered as removed from
the system (approximately for $p \geq 0.5$).  The 
percolation model introduced in Ref.~\cite{Havlin1}
instead describes a much more restrictive situation, where the largest
cluster is formed by airports that are connected simultaneously 
by all three carriers. 
The
considerations reported above are valid for results 
valid both for a single instance 
of the percolation model (Fig.~\ref{fig:fig2}a), and for 
average values over multiple instances of the model (Fig.~\ref{fig:fig2}b).
}

 {
As it appears clear from Figure~\ref{fig:fig2},
our theory is able to reproduce with high accuracy the results of
numerical simulations. The next sections are devoted to
the description of a complete mathematical
framework that allows us for the description of the redundant percolation
model. We stress that the framework is devised for
arbitrary topologies, and can be therefore applied safely
also to real multilayer networks as long as their structures are sufficiently
compatible with  the locally treelike approximation.
}

\section{Message-passing  algorithm}

We assume that 
interactions within each layer $\alpha$ are described by elements $a_{ij}^{[\alpha]}$
of the adjacency matrix of the layer, 
indicating whether the replica nodes 
$(i,\alpha)$ and $(j,\alpha)$ are connected ($a_{ij}^{[\alpha]}=1$) 
or not  ($a_{ij}^{[\alpha]}=0$) in layer $\alpha$.
Additionally, we consider a specific realization of the initial 
damage to the replica nodes indicated by the set $\{s_{i\alpha}\}$. The
generic element $s_{i\alpha}=0$ indicates that the 
replica node $(i,\alpha)$ has been initially damaged,  whereas 
$s_{i\alpha}=1$ indicates that the replica node $(i,\alpha)$ has not been initially damaged.
Under these conditions, as long as the  {multiplex} network is locally
treelike, the following  
message-passing algorithm identifies the 
replica nodes that are in  the RMCGC.

Each node $i$ sends to a 
neighbor $j$ a set of messages 
$n_{i\to j}^{[\alpha]}$ in every layer $\alpha$ where node $i$ is
connected to node $j$, i.e., 
with $a^{[\alpha]}_{ij}=1$.
These messages indicate 
whether ($n_{i\to j}^{[\alpha]}=1$) or 
not ($n_{i\to j}^{[\alpha]}=0$) node $i$
 connects node $j$ to the RMCGC with links 
belonging to layer $\alpha$.
The message 
$n_{i\to j}^{[\alpha]}=1$ if and only 
if all the following conditions are met:
\begin{itemize}
\item[(a)] node $i$ is connected to node $j$ in layer $\alpha$,
 and both nodes $(i,\alpha)$ and node $(j,\alpha)$ are not damaged, i.e., $s_{i\alpha}=s_{j\alpha}=a^{[\alpha]}_{ij}=1$;
\item[(b)] node $i$ is connected to the RMCGC through at least one node $\ell\neq j$ in layer $\alpha$;
\item[(c)] node $i$ belongs to the RMCGC assuming that  also node $j$ belongs to the RMCGC.
This conditions is satisfied if and only if,  assuming that node $j$
belongs 
to the RMCGC, node $i$ is connected in at least two layers to the RMCGC.
\end{itemize}
If the previous conditions are not simultaneously met, then $n_{i\to j}^{[\alpha]}=0$.
Put together, the former conditions lead to 
\bea
n_{i\to j}^{[\alpha]}=\theta(v_{i \to j},2) \,
a^{[\alpha]}_{ij}s_{j\alpha}s_{i\alpha}\left[1-\prod_{\ell\in
    N_{\alpha}(i)\setminus j}\left(1-n^{[\alpha]}_{\ell\to
      i}\right)\right] .
\label{m1}
\eea
Here $N_{\alpha}(i)$ indicates the set of nodes that are neighbors of 
node $i$ in layer $\alpha$. The term $1-\prod_{\ell\in
  N_{\alpha}(i)\setminus j}\left(1-n^{[\alpha]}_{\ell\to i}\right)$
therefore will equal one if at least one message is arriving to node
$i$
from a neighboring node $\ell \neq j$, while it will be equal to zero, otherwise.
$\theta(v_{i \to j},2)=1$ for $v_{i \to j}\geq 2$ and
$\theta(v_{i \to j},2)=0$, otherwise.
$v_{i \to j}$ indicates in how many 
layers node $i$ is connected to the RMCGC assuming that 
node $j$ also belongs to the RMCGC, i.e., 
\bea
v_{i\to j}&=&\sum_{\alpha=1}^M\left[s_{i\alpha}\left(1-\prod_{\ell\in N_{\alpha}(i)\setminus j}\left(1-n^{[\alpha]}_{\ell\to i}\right)\right)\right.\nonumber \\
&&\hspace{-2mm}\left.+s_{i\alpha}s_{j\alpha}a^{[\alpha]}_{ij}\prod_{\ell\in N_{\alpha}(i)\setminus j}\left(1-n^{[\alpha]}_{\ell\to i}\right)\right].
\label{v}
\eea
Therefore $v_{i\to j}$ indicates the  number of  initially undamaged
replica nodes $(i,\alpha)$ that either receive at least one positive
messages from nodes $\ell\in N_{\alpha}(i)\setminus j$ 
or are connected to the undamaged replica nodes $(j,\alpha)$.
Finally, the replica node $(i,\alpha)$ belongs to the RMCGC if (i) it
is not damaged, (ii) it is connected to the RMCGC in layer $\alpha$,
 and (iii) it  receives at least another positive message 
in a layer $\alpha'\neq \alpha$. 
These conditions are summarized by 
\bea
\sigma_{i\alpha}&=&s_{i\alpha}\left(1-\prod_{\ell \in N_{\alpha}(i)}\left(1-n^{[\alpha]}_{\ell\to i}\right)\right)\nonumber \\
&&\hspace{-15mm}\times \left\{1-\prod_{\alpha' \neq \alpha}\left[1-s_{i\alpha'}+s_{i\alpha'}\prod_{\ell \in N_{\alpha'}(i)}\left(1-n^{[\alpha']}_{\ell\to i}\right)\right]\right\}.
\label{s1}
\eea
The average number $S$ 
of replica nodes belonging to the RMCGC is computed as
\bea
S=\frac{1}{MN}\sum_{\alpha=1}^M\sum_{i=1}^N \sigma_{i\alpha}.
\label{S}
\eea
The system of Eqs.~(\ref{m1}),~(\ref{v}),~(\ref{s1}), and~(\ref{S})
represents a complete mathematical framework to estimate
the average size of the RMCGC for a given 
network and a given initial configuration
of damage. The solution can be obtained by first iterating 
Eqs.~(\ref{m1}) and~(\ref{v}) to obtain the values of the messages
$n_{i \to j}^{[\alpha]}$. Those values are then plugged into
Eqs.~(\ref{s1}) to compute the values of the variables $s_{i \alpha}$,
and finally these variables are used into Eq.~(\ref{S}) to estimate
the average size of the RMCGC. We stress that, being valid for a given
network and for a given configuration of damage, the values of
the variables $n_{i \to j}^{[\alpha]}$ and $s_{i \alpha}$ are either $0$ or $1$.
The variables $v_{i \to j}$ can assume instead integer values in the
range $[0, M]$. 
The mathematical framework works properly
also in presence of edge overlap among
layers. This is an important feature that can change 
dramatically change the robustness 
properties of multilayer networks~\cite{Goh_comment, Cellai2013,
  Radicchi,Baxter2016, Cellai2016}.

\section{ Multiplex networks without link overlap}
\subsection{General results}
\subsubsection{Simplification of the message-passing equations on a single realization of the initial damage}
In absence of link overlap, 
a given pair of nodes 
$i$ and $j$ may be linked 
exclusively along a 
single layer $\alpha$. 
Nontrivial messages potentially
different from zero will therefore exist
only on a specific layer for every pair
of connected nodes $i$ and $j$. 
It can be easily seen that the 
message-passing Eqs.~(\ref{m1}) and~(\ref{v}) reduce to   
\bea
n_{i\to j}^{[\alpha]}&=&s_{i\alpha}s_{j\alpha}a_{ij}^{[\alpha]}\left[1-\prod_{\ell \in N_{\alpha}(i)\setminus j}\left(1-n_{\ell\to i}^{[\alpha]}\right)\right]\nonumber \\
&&\hspace*{-17mm}\times \left\{1-\prod_{\alpha'\neq \alpha}\left[1-s_{i\alpha'}+s_{i\alpha'}\prod_{\ell \in N_{\alpha'}(i)}\left(1-n_{\ell\to i}^{[\alpha']}\right)\right]\right\}.
\label{m2}
\eea

We further notice that in this situation the result of the
message-passing 
algorithm does not change if we consider messages that depend
exclusively on the state $s_{i\alpha}$ of the node $i$ that sends the
message.  Even if we drop the factor $s_{j\alpha}$ in
Eq.~(\ref{m2}),  the message will be allowed anyways 
to propagate further at the next
iteration step,  if the replica node $(j,\alpha) $ is not
initially damaged. 
Therefore, we can further simplify Eq.~(\ref{m2})
and consider
\bea
n_{i\to j}^{[\alpha]}&=&s_{i\alpha}a_{ij}^{[\alpha]}\left[1-\prod_{\ell \in N_{\alpha}(i)\setminus j}\left(1-n_{\ell\to i}^{[\alpha]}\right)\right]\nonumber \\
&&\hspace*{-17mm}\times \left\{1-\prod_{\alpha'\neq \alpha}\left[1-s_{i\alpha'}+s_{i\alpha'}\prod_{\ell \in N_{\alpha'}(i)}\left(1-n_{\ell\to i}^{[\alpha']}\right)\right]\right\}.
\label{m3}
\eea
Eqs.~(\ref{m3}) replace Eqs.~(\ref{m1}) and~(\ref{v}) in the case of a
multiplex
network without link overlap. The rest of the framework is identical,
so that Eqs.~(\ref{s1}) and~(\ref{S}) 
remain unchanged.

\subsubsection{Message-passing equations for random realizations of
  the initial damage}
Eqs.~(\ref{m3}),~(\ref{s1}), and~(\ref{S}) 
determine the average size of the RMCGC in a multiplex network
without link overlap for a
given realization of the initial damage $\{s_{i\alpha}\}$.
These equations  can be, however, extended to make predictions 
in the case of a random realization of the initial damage 
when the replica nodes are damaged independently with probability
$1-p$, i.e,. such that the 
the initial damage $\{s_{i\alpha}\}$
is a random configuration 
obeying the probability distribution
\bea
\hat{\mathcal P}( \{s_{i\alpha}\})=\prod_{i=1}^N\prod_{\alpha=1}^M p^{s_{i\alpha}}(1-p)^{1-s_{i\alpha}}.
\label{ps}
\eea
To this end, we denote the probability that node $i$ sends
a positive message to node $j$ in layer $\alpha$ 
by  
$\hat{n}_{i\to j}^{[\alpha]}=\Avg{n_{i\to j}^{[\alpha]}}$, and 
the probability that the replica node $(i, \alpha)$ belongs 
to the RMCGC 
by $\hat{\sigma}_{i\alpha}=\Avg{\sigma_{i\alpha}}$.
The message-passing algorithm 
determining the values of $\hat{n}_{i\to j}^{[\alpha]}$ and $\hat{\sigma}_{i\alpha}$ is given by 
\bea
\hat{n}_{i\to j}^{[\alpha]}&=&a_{ij}^{[\alpha]}p\left[1-\prod_{\ell \in N_{\alpha}(i)\setminus j}\left(1-\hat{n}_{\ell\to i}^{[\alpha]}\right)\right]\nonumber \\
&&\hspace*{-17mm}\times \left\{1-\prod_{\alpha'\neq \alpha}\left[1-p+p\prod_{\ell \in N_{\alpha'}(i)}\left(1-\hat{n}_{\ell\to i}^{[\alpha']}\right)\right]\right\},\nonumber \\
\hat{\sigma}_{i\alpha}&=&p\left(1-\prod_{\ell \in N_{\alpha}(i)}\left(1-\hat{n}^{[\alpha]}_{\ell\to i}\right)\right)\nonumber \\
&&\hspace{-17mm}\times \left\{1-\prod_{\alpha' \neq \alpha}\left[1-p+p\prod_{\ell \in N_{\alpha'}(i)}\left(1-\hat{n}^{[\alpha']}_{\ell\to i}\right)\right]\right\}.
\eea

This algorithm can be applied to 
a given network, and provides the 
average number of replica nodes $S$ belonging to the RMCGC for a
random realization of the initial damage obeying 
Eq. $(\ref{ps})$. Specifically the value of $\hat{S}$ is related to $\hat{\sigma}_{i\alpha}$ by 
\bea
\hat{S}=\frac{1}{MN}\sum_{i=1}^N \sum_{\alpha=1}^M \hat{\sigma}_{i\alpha}.
\eea

\subsubsection{Message-passing equations for random multiplex
  networks}
A  {multiplex} network where every layer is a sparse 
network generated according to the
configuration model
is a major example of a multiplex network 
without link overlap in the limit of large network sizes.
It is therefore natural and important to characterize 
the RMCGC in this case.
We assume that every network layer
$G_{\alpha}$ is a random graph
taken from the probability distribution 
\bea
{\mathcal P}^{[\alpha]}(G_{\alpha})=\frac{1}{Z}\prod_{i=1}^N \delta \left(k_i^{[\alpha]},\sum_{j=1}^N a_{ij}^{[\alpha]} \right),
\label{ens}
\eea
where  $k_i^{[\alpha]}$ indicates the pre-imposed degree of node $i$ 
in  layer $\alpha$, $\delta(x,y) =1$ if $x=y$ and $\delta(x,y)=0$,
otherwise,
and $Z$ is the normalization factor 
indicating the total number of networks in the ensemble.
Averaging over the network ensemble 
allows us to translate the message-passing equations 
into simpler expressions for the characterization 
of the percolation transition. 
 
Let us consider a random \change{multiplex} network obeying the probability 
of Eq.~(\ref{ens}),  and a random realization of the initial 
damage described by the probability of Eq.~(\ref{ps}). The 
average message in layer 
$\alpha$, namely $S'_{\alpha}=\Avg{\hat{n}_{i\to
    j}^{\alpha}|a_{ij}^{[\alpha]}=1}$, 
and the  average number of replica nodes of layer $\alpha $ 
that are in the RMCGC,  denoted by
$S_{\alpha}=\Avg{\hat{\sigma}_{i,\alpha}}$, obey the equations
\bea
S_{\alpha}&=&p\sum_{\bf k}P({\bf k})\left[1-(1-S'_{\alpha})^{k^{[\alpha]}}\right]\nonumber \\
&&\left\{1-\prod_{\alpha'\neq \alpha}\left[1-p+p(1-S'_{\alpha'})^{k^{[\alpha']}}\right]\right\},
\nonumber 
\\
S'_{\alpha}&=&p\sum_{\bf k}\frac{k^{\alpha}}{\Avg{k^{\alpha}}}P({\bf k})\left[1-(1-S'_{\alpha})^{k^{[\alpha]}-1}\right],
\nonumber 
\\
&&\times\left\{1-\prod_{\alpha'\neq \alpha}\left[1-p+p(1-S'_{\alpha'})^{k^{[\alpha']}}\right]\right\},
\label{SU1}
\eea
where $P({\bf k})$ indicates the probability that a generic node $i$ has degrees ${\bf k}_i={\bf k}$, i.e. $(k^{[1]}_i,k^{[2]}_i,\ldots, k^{[M]}_i)=(k^{[1]},k^{[2]},\ldots, k^{[M]})$.

If there are no correlations between the degrees of a node in
different layers,  the degree distribution $P({\bf k})$ can be factorized
as 
\bea
P({\bf k})=\prod_{\alpha}P^{[\alpha]}(k^{[\alpha]}) \; , 
\label{eq:pk}
\eea
where $P^{[\alpha]}(k)$ is the degree distribution in layer $\alpha$.
In this case, Eqs. $(\ref{SU1})$ can be expressed in terms 
of the generating function of the degree distribution in each layer. Specifically, we have 
\bea
S_{\alpha}&=&p\left[1-H_0^{[\alpha]}(1-S'_{\alpha})\right]\nonumber \\
&&\left\{1-\prod_{\alpha'\neq \alpha}\left[1-p+p \ H_0(1-S'_{\alpha'})\right]\right\},
\nonumber 
\\
S'_{\alpha}&=&p\left[1-H_1^{[\alpha]}(1-S'_{\alpha})\right]
\nonumber \\
&&\left\{1-\prod_{\alpha'\neq \alpha}\left[1-p+p\ H_0^{[\alpha']}(1-S'_{\alpha'})\right]\right\},
\label{SU2}
\eea
where the  generating functions $H_0^{[\alpha]}(z)$
and $H_1^{[\alpha]}(z)$ of the degree distribution $P^{[\alpha]}(k)$ of layer $\alpha$  are given by 
\bea
H_0^{[\alpha]}(x)&=&\sum_{k}P^{[\alpha]}(k) \, x^k, \nonumber \\
H_1^{[\alpha]}(x)&=&\sum_{k}\frac{k}{\avg{k^{[\alpha]}}}P^{[\alpha]}(k)
\, x^{k-1}.\label{SLMCGC}
\eea

Finally the average number $S$ of replica nodes in the RMCGC is given by 
\bea
S=\frac{1}{M}\sum_{\alpha} S_{\alpha}.
\eea
 
If we consider the
case of equally distributed Poisson layers with average degree $z$, we
have 
that Eq.~(\ref{eq:pk}) is
\bea
P^{[\alpha]}(k)=\frac{1}{k!}z^ke^{-z}
\eea
for every layer $\alpha=1,2,\ldots, M$.
Then, using Eqs.~(\ref{SU2}), 
one can show that $S'_{\alpha}=S_{\alpha}=S$, $\forall \alpha$, 
and $S$ is determined by the equation 
\bea
S=p\left(1-e^{-zS}\right)\left\{1-[1-p+pe^{-zS}]^{M-1}\right\}.
\label{P1}
\eea
This equation has always the trivial solution $S=0$. In addition, a nontrivial solution $S>0$ indicating the presence of the RMCGC, emerges at a hybrid discontinuous transition  characterized by a square root singularity, on a line of points $p=p_c(z)$,  determined by the equations 
\bea
h_{z,p}(S_c)&=&0,\nonumber \\
\left.\frac{dh_{z,p}(S)}{dS}\right|_{S=S_c}&=&0 ,
\eea 
where
\bea
h_{z,p}(S)&=&S-p(1-e^{-zS})\nonumber \\
&&\times \left\{1-[1-p+pe^{-zS}]^{M-1}\right\}=0.
\eea
For $p> p_c$ there is a RMCGC, for $p \leq p_c$ there is no RMCGC. The entity of the discontinuous jump at $p=p_c$ in the fraction $S$ of replica nodes  in the RMCGC is given by $S=S_c$.
The percolation threshold  $p_c$ as a function of the average degree $z$ of the network is plotted in Figure $\ref{fig:pcPoisson}$ for $M=2,3,4,5$.
It is shown that as the number of layers $M$ increases the percolation threshold decreases for every value of the average degree $z$.
Also, the discontinuous jump $S_c$ decreases as the number $M$ of layer increases  for very given average degree $z$ (see Figure $\ref{fig:ScPoisson}$).
Therefore, as the number of layers increases the multilayer networks becomes more robust.

\begin{figure}[htb]
\begin{center}
	\includegraphics[width=1.0\columnwidth]{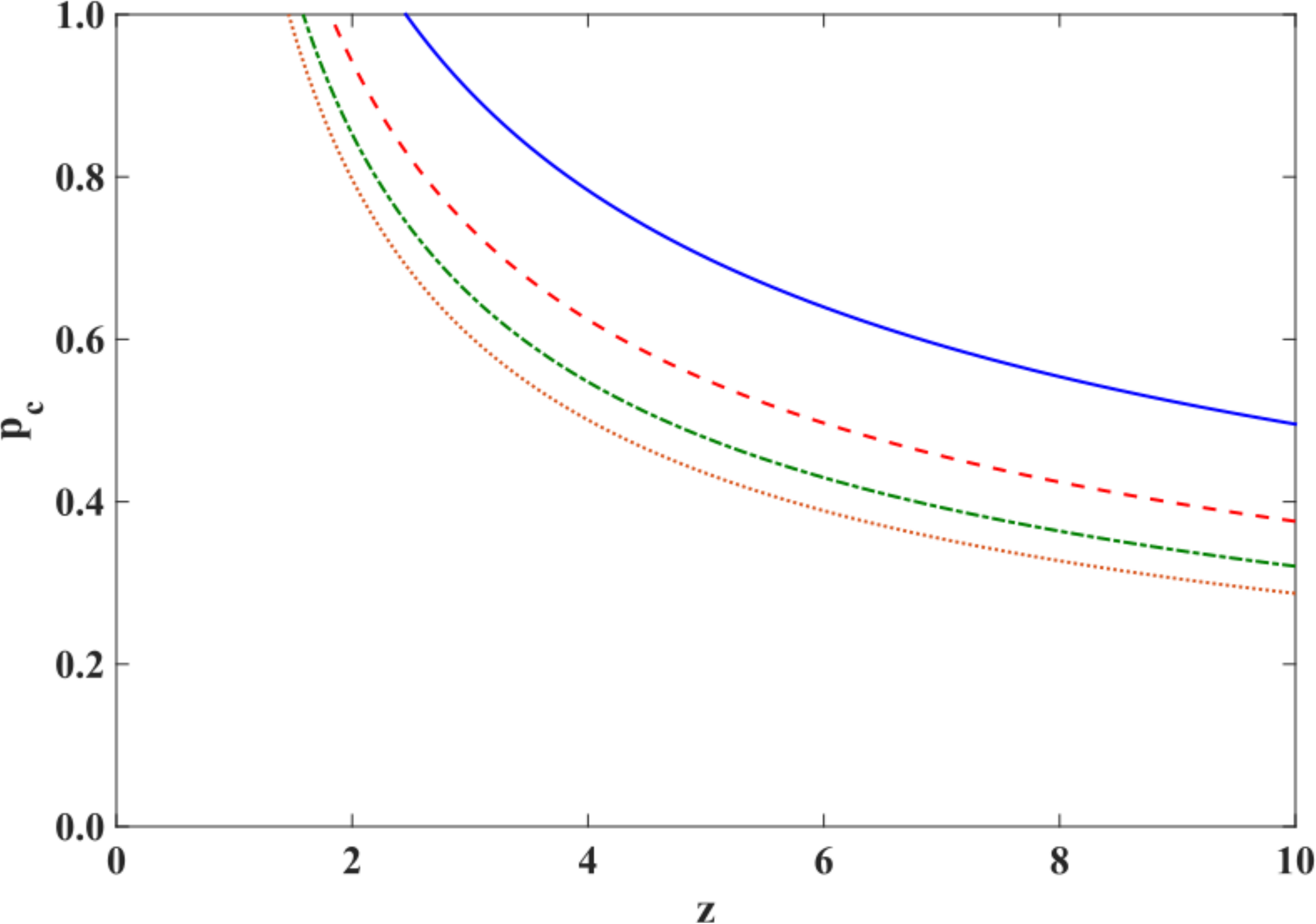}
	\caption{
The percolation threshold $p_c$ is plotted versus the average degree $z$ of each layer for  Poisson multiplex networks with $M=2,3,4,5$ layers indicated respectively with with blue solid, red dashed, green dot-dashed and orange dotted lines.}	
	\label{fig:pcPoisson}
\end{center}
\end{figure}

\begin{figure}[htb]
\begin{center}
	\includegraphics[width=1.0\columnwidth]{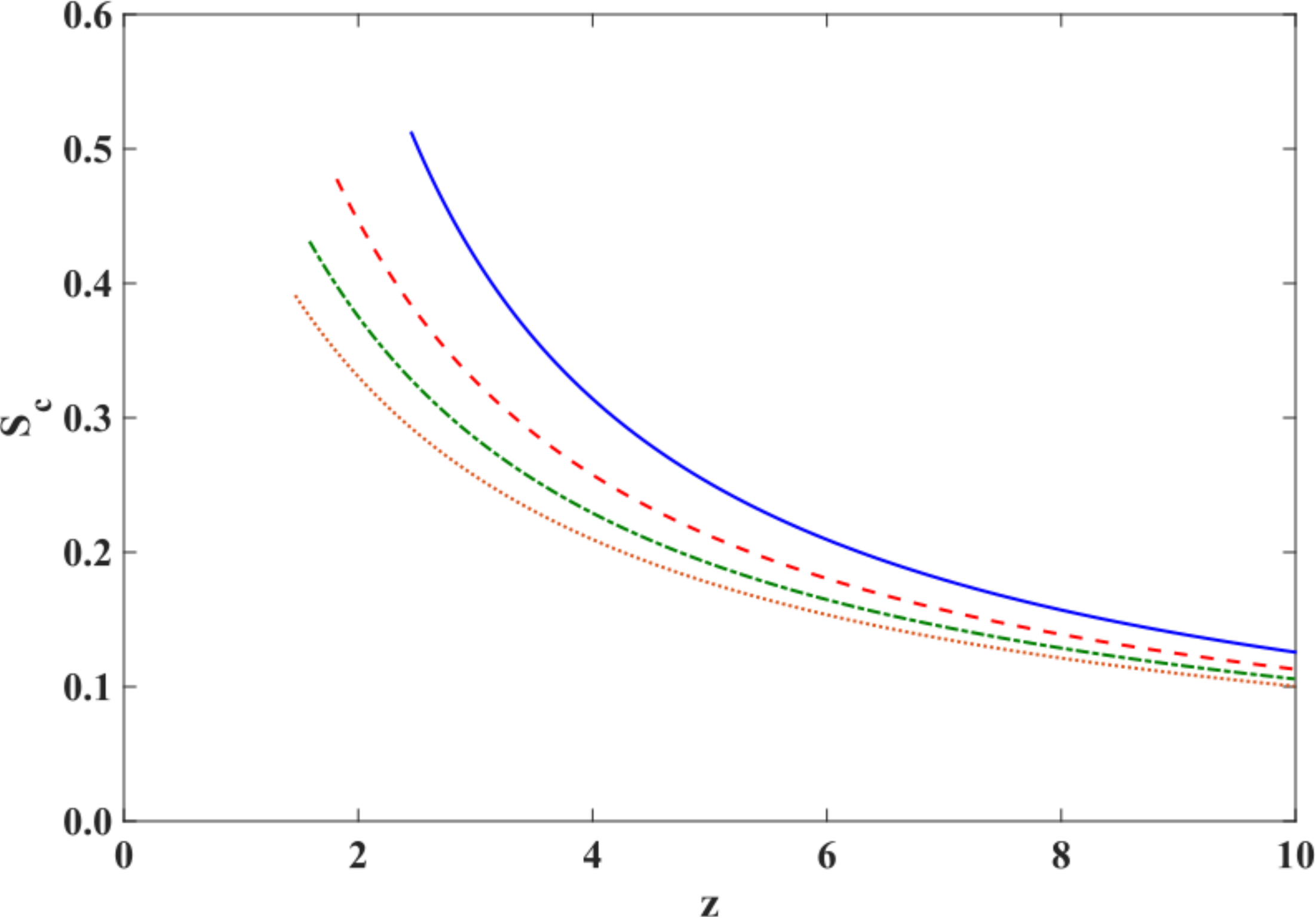}
	\caption{The discontinuous jump  $S_c = S(p_c)$ of the RMCGC at the
          percolation threshold $p=p_c$,  is plotted versus the
          average degree $z$ of each layer for  Poisson multiplex
          networks with $M=2,3,4,5$ 
         layers indicated respectively with blue solid, red dashed, 
         green dot-dashed and orange dotted lines.}	
	\label{fig:ScPoisson}
\end{center}
\end{figure}
\subsection{Comparison between the RMCGC and the MCGC }
In this section, we compare the robustness of \change{multiplex }networks in presence of ordinary interdependencies and in presence of redundant interdependencies.
To take a concrete example, we consider the case of a multiplex 
network with $M$ Poisson layers, each layer having the same average degree $z$.
In this case the fraction $S$ of replica nodes in the RMCGC is given
by the solution of Eqs.~(\ref{P1}) 
while the fraction of replica nodes in the MCGC is given by~\cite{gao2011}
\bea
S=\tilde{p}\left(1-e^{-zS}\right)^M.
\label{o}
\eea
In Eq.~(\ref{o}),  it is  assumed that  every replica node $(i,\alpha)$ of a given node $i$ is damaged  simultaneously (with probability $\tilde{f}=1-\tilde{p}$). 
On the contrary,  in presence of redundant interdependencies it is
natural
to assume that the initial damage is inflicted to each replica node independently (with probability $f=1-p$).
Therefore, in order to compare the robustness of the multiplex
networks in presence and in absence of redundant interdependencies, we
set $p=\tilde{p}=1$, i.e., replica nodes are not initially damaged,
and 
compare the critical value of the average degree $z=z^{\star}$ at 
which the percolation transition occurs respectively for the RMCGC and for the MCGC.
Additionally we will characterize also the size $S=S^{\star}$ of the
jump in the 
size of the RMCGC and the MCGC at the percolation transition.
In Fig.~\ref{fig:Comp_M}, we display the values of $z^{\star}$ and
$S^{\star}$ 
as a function of the number of layers $M$ for the RMCGC and the 
MCGC. For $M=2$, the two models give the same results as they are identical.
For $M > 2$, differences arise. In presence of redundant
interdependencies, 
multiplex networks
become increasingly more robust as the number $M$ of layers
increases. This phenomenon is apparent from the fact that the RMCGC
emerges for multiplex networks 
with an average degree of their layers $z^{\star}$ which decreases as the number of layers $M$ increases.
On the contrary, in ordinary percolation the value of $z^{\star}$ 
for the emergence of the MCGC is an increasing function of $M$.
Additionally, the size of the discontinuous jumps $S^{\star}$ at the
transition point decreases with $M$ for the RMCGC, while increases with
$M$ for the MCGC showing that the avalanches of failures have a
reduced size for  the RMCGC.
\change{We expect that the  beneficial effect of the
  addition of new layers  will extend also to the scenario in which
  the number of operating layers necessary for a node to
be functional is assumed
to be  lager than two. Also, we believe that the
same conclusion will apply to more general multilayer network
structures where nodes  have different number of interlinks (redundant
interdependencies) such as a the topologies considered in
Ref.~\cite{BD2} as long as the number of operating layers necessary
for a node to be functional is the same for every node.}
\begin{figure}[htb]
\begin{center}
	\includegraphics[width=1.\columnwidth]{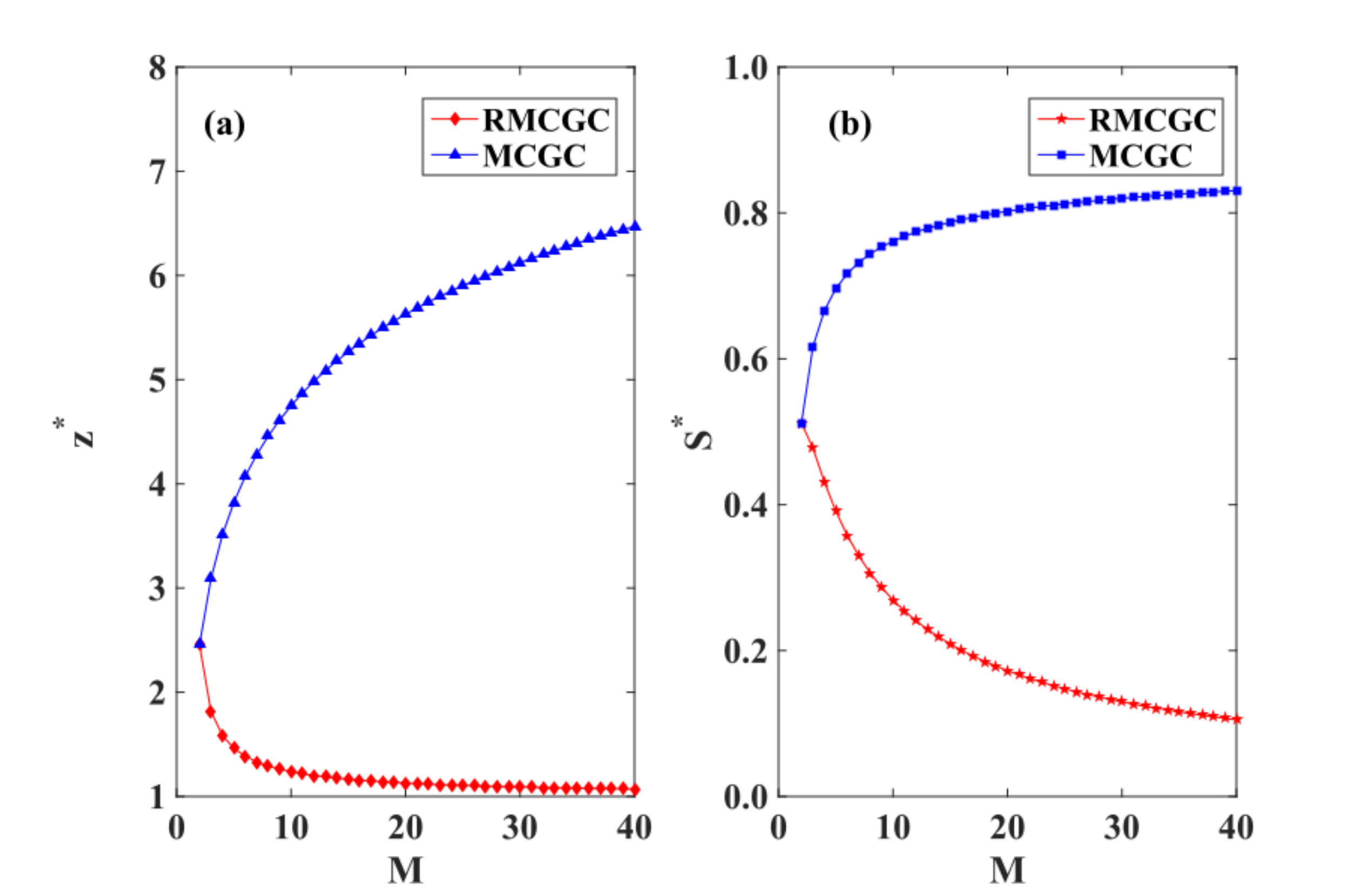}	
	\caption{
          Comparison between the MCGC and the RMCGC models
        in Poisson multiplex networks. (a) Critical value $z^{\star}$ of
        the average degree as a function of the number of network
        layers $M$. Results
        for the RMCGC model are displayed as red diamonds. Results for 
        the MCGC model are denoted by blue triangles. (b) Height of
        the jump $S^{\star}$ at the transition
        point as a function of the number of network layers. 
      }
	\label{fig:Comp_M}
\end{center}
\end{figure}

\subsection{Comparison with numerical simulations}
\label{Ss1}
In this section, we compare the results 
obtained with Eqs.~(\ref{m1}),~(\ref{v}),~(\ref{s1}), and~(\ref{S})
on a single instance of damage with the predictions the message-passing
algorithm described 
in Eq.~(\ref{SU2}) characterizing the size $S$ of the RMCGC in an 
ensemble of networks.
Specifically, we consider the case of a multilayer network with $M=3$ Poisson layers with the same average degree $z$.
In order to draw the percolation diagram for single instances of
initial damage as a function of the probability of damage $1-p$,
we  associate each replica node $(i,\alpha)$ with a random variable
$r_{i\alpha}$ drawn from a uniform distribution and we set 
\bea
s_{i\alpha}=\left\{\begin{array}{ccc} 1&\mbox{if}& r_{i\alpha}\leq p\\
0&\mbox{if}& r_{i\alpha}>p \end{array}\right.
\label{rs}
\eea
Fig.~\ref{fig:MP_single} displays the comparison between the two
approaches, showing an almost perfect agreement between them.
\begin{figure}[htb]
\begin{center}
	\includegraphics[width=1\columnwidth]{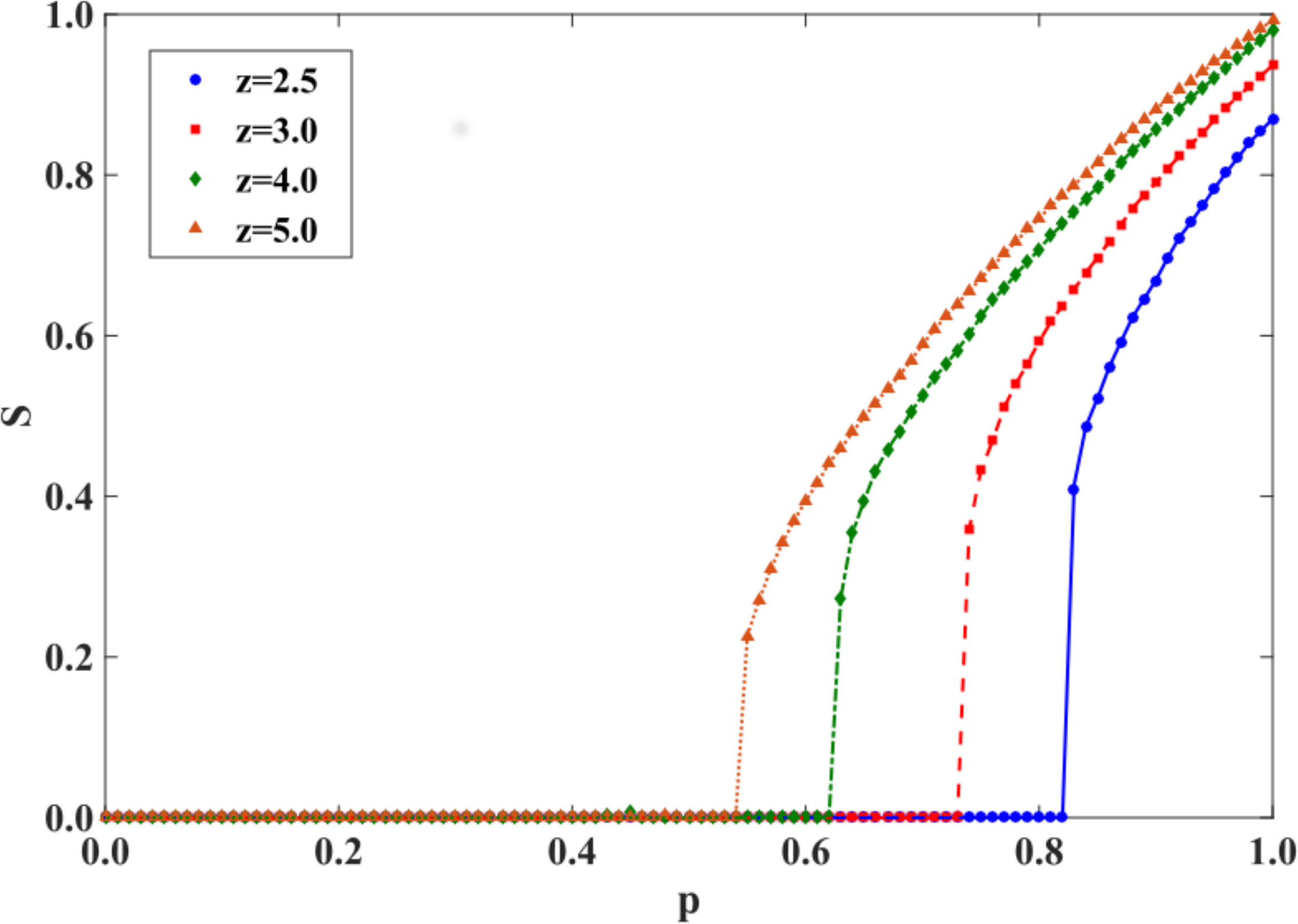}
	\caption{Comparison between simulation results of the
          RMCGC for a multiplex network with $M=3$ Poisson layers
          Poisson with average degree $z$ and no link overlap,  
         and the message-passing results over single network
         realization and given configuration damage.
         We consider different values of the average degree $z=2.5,
         3.0, 4.0 , 5.0$.
         Points indicate results of numerical simulations: blue
         circles ($z=2.5$), red squares ($z=3.0$), green diamonds ($z=4.0$), and
         orange triangles ($z=5.0$). 
        Message-passing  
        predictions are denoted by lines with the same color scheme
        used for numerical simulations. 
        Simulations results are performed on a single instance of a
        multiplex network 
        with $N=10^4$ nodes.}	
	\label{fig:MP_single}
\end{center}
\end{figure}
Additionally in Fig.~\ref{fig:single_average}, we compare simulation
results averaged over several 
realizations of the initial damage  and several instances of  the
multiplex network model  with the theoretical predictions given by
the numerical solution of  Eqs.~(\ref{SU1})-(\ref{P1}),
obtaining a very good agreement.

\begin{figure}[htb]
\begin{center}
	\includegraphics[width=1\columnwidth]{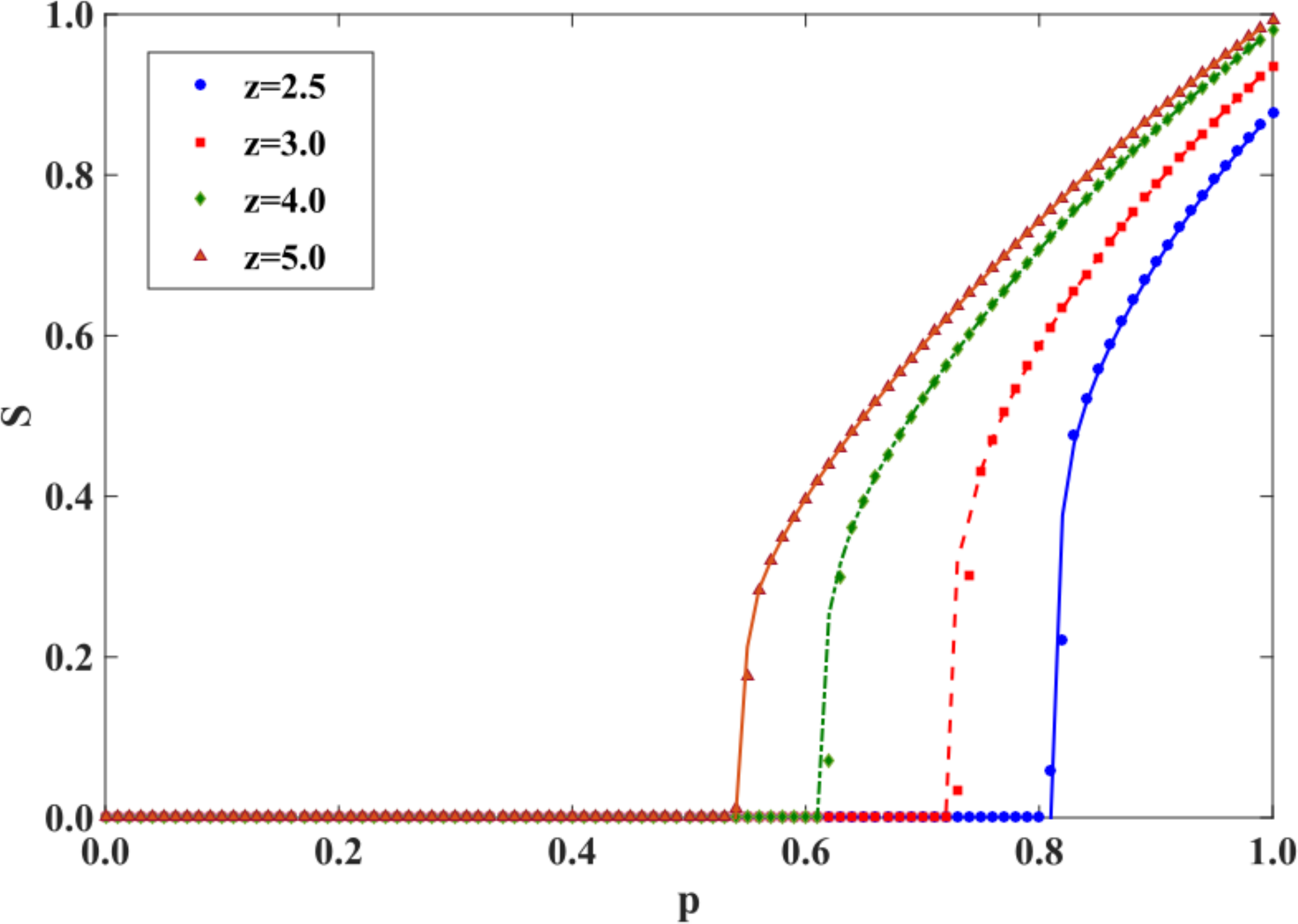}
	\caption{Same as in Fig.~\ref{fig:MP_single}, but for averages
          over $20$
          instances of the multiplex network model and configurations
          of random initial damage.}	
	\label{fig:single_average}
\end{center}
\end{figure}
\section{MULTIPLEX NETWORKS WITH LINK OVERLAP}

\subsection{Link overlap, multilinks and multidegree}

In isolated networks, two nodes can be either connected or not
connected. 
In multiplex networks instead, the complexity of the structure 
greatly increases as the ways in which a generic pair of
nodes can be connected is given by $2^M$ possibilities. 
A very convenient way of accounting for all the possibilities with a
compact notation is to use the notion of multilink among pairs of
nodes~\cite{PRE,Weighted}.
Multilinks $\vec{m}=\left(m^{[1]},m^{[2]},\ldots, m^{[M]}\right)$ 
with $m^{[\alpha]}=0,1$,  describe any  of the possible patterns of
connections 
between pairs of nodes in a multilayer network with $M$ layers. 
Specifically, $m^{[\alpha]}=1$ indicates that a connection exists in
layer 
$\alpha$, whereas $m^{[\alpha]}=0$ indicates that 
the  connection in layer $\alpha$ does not exists.
In particular, we can say that, in a multiplex network with $M$
layers, two nodes $i$ and $j$  are connected 
by the multilink
\bea
\vec{m}_{ij}=(a_{ij}^{[1]},a_{ij}^{[2]},\ldots, a_{ij}^{[M]}).
\eea

In order to distinguish the case in which two nodes are not connected
in any layer with the case in which in at least one layer the nodes
are connected, we distinguish between  the trivial multilink 
$\vec{m}=\vec{0}$ and the nontrivial multilinks 
$\vec{m}\neq \vec{0}$. The trivial multilink 
$\vec{m}=\vec{0}$ indicates the absence of any 
sort of link between the two nodes.

Using the concept  of multilinks,  one can define  
multiadjacency matrices ${\bf A}^{\vec{m}}$ whose element
$A_{ij}^{\vec{m}}$ 
indicates whether ($A_{ij}^{\vec{m}}=1$)  or not
$(A_{ij}^{\vec{m}}=0)$ a  
node $i$ is connected to node $j$ by a multilink $\vec{m}$.
The matrix elements $A_{ij}^{\vec{m}}$ of the multiadjacency matrix
${\bf A}^{\vec{m}}$ 
are given by  
 \bea
 A_{ij}^{\vec{m}}=\prod_{\alpha=1}^M\delta\left(m^{[\alpha]},a_{ij}^{[\alpha]}\right).
 \eea
 {We note that multiadjacency matrices
are essentially equivalent to rank-3 tensors as those 
considered in Ref.~\cite{Mucha} for multiplex networks,
and generalized to the
case of arbitrary multilayer networks in Ref.~\cite{Math}. 
}
 Using multiadjacency matrices, it is straightforward
 to define 
multidegrees \cite{PRE,Weighted}. 
 The  multidegree 
of node $i$ indicated as $k_i^{\vec{m}}$ 
is the sum of rows (or columns)
 of the multiadjacency matrix ${\bf A}^{\vec{m}}$, i.e.,
 \bea
 k_i^{\vec{m}}=\sum_j A_{ij}^{\vec{m}},
 \eea
 and indicates how many multilinks $\vec{m}$ are incident to node $i$.

 Using a multidegree sequence $\{k_i^{\vec{m}}\}$, it is possible to
 build multiplex network 
ensembles that generalize the configuration model.
This way, overlap of links is fully preserved by the randomization 
of the multiplex network. 
These ensembles are specified by the 
probability $\tilde{\mathcal P}(\vec{G})$ attributed
 to every multiplex network $\vec{G}$ of the ensembles, 
where $\tilde{\mathcal P}(\vec{G})$ is given by 
\bea
\tilde{\mathcal P}(\vec{G})=\frac{1}{\tilde{Z}}\prod_{i=1}^N \prod_{\vec{m}\neq \vec{0}}\delta\left(k_i^{\vec{m}},\sum_{j=1}^N A_{ij}^{\vec{m}}\right),
\eea
 with $\tilde{Z}$ normalization constant equal to the number of
 multiplex networks 
with given multidegree sequence.

\subsection{General discussion of the message-passing equations for the RMCGC}

Our goal here is to  generalize the message-passing algorithm 
already given by Eqs.~(\ref{m1}),~(\ref{v}),~(\ref{s1}), and~(\ref{S})
for a generic single instance of a multiplex network and single
realization
of initial damage to the cases of (i) random multiplex networks with
given 
multidegree sequence and/or (ii) random realizations of the initial
damage.
The extensions for both cases has been already considered for the case
of
multiplex networks without link overlap.
In presence of link overlap, however, a more  {complex formalism is needed.}
For two nodes $i$ and $j$ in fact, the messages $n_{i\to
  j}^{[\alpha]}$ given by Eq.~(\ref{m1})  
and sent from node $i$ to node $j$ over the different layers
$\alpha=1,2\ldots M$ 
are correlated because they all depend on the value of the variable
$v_{i \to j}$ given by Eq.~(\ref{v}).
Such correlations require
particular care 
when averaging the messages to treat the percolation transition 
for random initial damages.
Similar technical challenges are also present in the treatment of the 
MCGC model where interdependencies are not redundant~\cite{Baxter2016,Cellai2016}.
In presence of redundant interdependencies there is an additional 
precaution that needs to be taken. In fact, the messages $n_{i\to
  j}^{[\alpha]}$  are 
explicitly dependent on the state of all replicas 
$(j,\alpha')$ of node $j$. This state is 
indicated by the variables $\vec{s}_j=(s_{j1}, s_{j2}, \ldots,
s_{j\alpha'}, \ldots s_{jM})$ 
where $s_{j\alpha'}$ specifies whether the replica node $(j,\alpha')$ is initially damaged or not.
As a consequence of this property, when averaging over random
realizations of initial damage, 
message-passing equations are written in terms of the messages 
$\hat{\sigma}_{i\to j}^{\vec{m}_{ij},\vec{n}}(\vec{s}_j)$ explicitly
accounting for the probability that node $i$ is sending to node $j$
the 
set of messages $\vec{n}=(n_{i\to j}^{[1]},n_{i\to j}^{[2]}\ldots
n_{i\to j}^{[\alpha]}, \ldots n_{i\to j}^{[M]})$,
given that node $j$ is in state $\vec{s}_j$ and node $i$ and node $j$ 
are connected by a multilink $\vec{m}=\vec{m}_{ij}$.
We have derived these equations for a general multiplex network with
$M$ layers. 
However, the message-passing algorithm has a very
long expression. 
To make the paper more readable, we decided to place the
exact treatment
of the general case in the SM, and consider here only the
special case of ensembles of random multilayer networks with overlap.
For these ensembles in fact, the message-passing equations are written in
terms of average messages 
sent between nodes with given multilinks 
$\vec{m}$, i.e., $S^{\vec{m},\vec{n}}(\vec{s}_j)=\Avg{\hat{\sigma}_{i\to j}^{\vec{m}_{ij},\vec{n}}(\vec{s}_j)|\vec{m}_{ij}=\vec{m}}$,
and the equations  greatly  simplify.  Two specific cases of multilayer
network ensembles are discussed below, for the cases of $M=2$ and $M=3$ layers.

\subsection{Ensembles of multilayer networks link overlap and $M=2$ layers}

In this case, every replica node is in
the RMCGC if and only if also 
its interdependent node in the other layer is 
in the RMCGC. Therefore, the only 
messages that are different from zero 
are the 
messages $S^{\vec{m},\vec{n}}(\vec{s}_j=(1,1))$ 
sent to nodes $j$  
in state $\vec{s}_j=(1,1)$. Specifically,  
we  consider  the case of a random multiplex network with
Poisson multidegree distributions characterized by the
averages 
\bea
\avg{k^{(1,1)}}&=&z_2,\nonumber \\
\avg{k^{(0,1)}}&=&\avg{k^{(1,0)}}=z_1.
\eea

The messages $S^{\vec{m},\vec{n}}(\vec{s}_j=(1,1))$ only depend on 
the multiplicity of overlap of the multilinks $\vec{m}$ 
and $\vec{n}$ given respectively by  
\bea
\mu &=& \sum_{\alpha=1}^M m^{[\alpha]},\nonumber \\
\nu &=& \sum_{\alpha=1}^M n^{[\alpha]}.
\eea
The  fraction $S$ of replica nodes in the RMCGC is determined by the variables
\bea
x_{\mu,\nu}=S^{\vec{m},\vec{n}}(\vec{s}_j=(1,1)),
\eea 
where for  example the value of  $x_{2,2}$ indicates  the probability  that node $i$ to sends a message $\vec{n}=(1,1)$ to its neighbor $j$ with $\vec{s}_j=(1,1)$ connected by a multilink $\vec{m}=(1,1)$. 

The values of the variables $x_{\mu,\nu}$ and   $S$ are   
determined by the following set of equations (See Supplementary Material for details)
\begin{equation}
\begin{array}{ll}
x_{2,2}= & p^2\left[1-2e^{-z_1x_{1,1}-z_2(x_{2,2}+x_{2,1})}
\right. \\
& \left. +e^{-2z_1x_{1,1}-z_2(x_{2,2}+2x_{2,1})}\right]  \\
x_{2,1}= & p^2\left[e^{-z_1x_{1,1}-z_2(x_{2,2} +x_{2,1})}\right.  \\
& \left. -e^{-2z_1x_{1,1}-z_2(x_{2,2}+2x_{2,1})}\right] \\
S= & x_{1,1}=x_{2,2}
\end{array} \;.
\end{equation}

These equations are the same equations as those that determine the 
value of the MCGC as long we make the substitution $p^2\to p$~\cite{Baxter2016,Cellai2016} taking into account 
 that the damage in each replica node is  independent in the present model.  {Notably in this case the percolation phase transition is discontinuous and hybrid  been characterized by a square-root singularity above for $p$ approaching the percolation threshold $p_c$ from above.}

\subsection{Ensembles of multilayer networks link overlap and $M=3$ layers}

\begin{figure*}[htb]
\begin{center}
	\includegraphics[width=1.5\columnwidth]{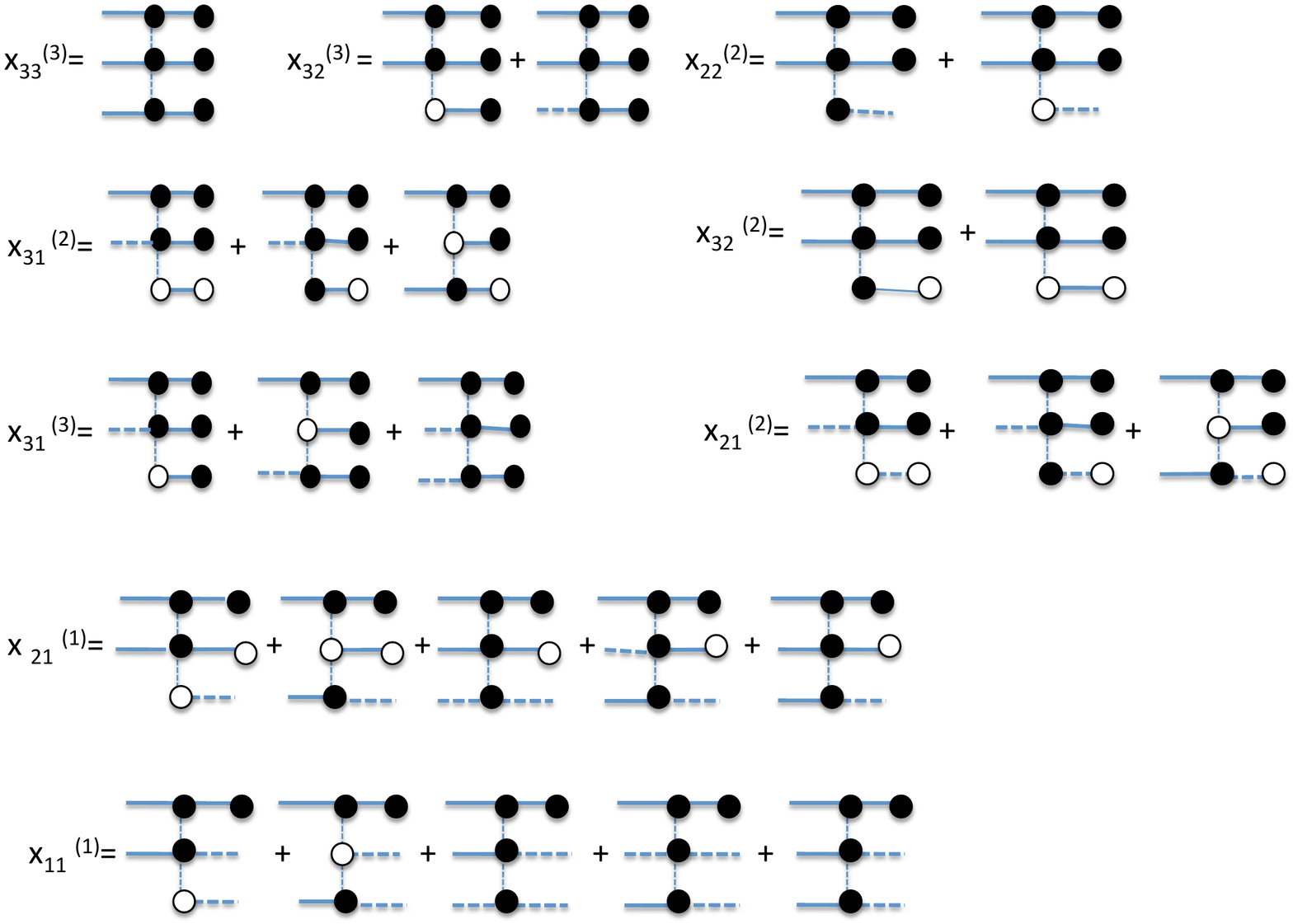}
	\caption{Diagrams for Eqs. $(\ref{m_3l})$ determining $x_{\mu
            \nu}^{(\xi)}$ in the case of multiplex networks with three layers ($M=3$) and Poisson multidegree distribution with $\avg{k^{(1,1,1)}}=z_3$, $\avg{k^{(1,1,0)}}=\avg{k^{(1,0,1)}}=\avg{k^{(0,1,1)}}=z_2$ and $\avg{k^{(1,0,0)}}=\avg{k^{(0,1,0)}}=\avg{k^{(0,0,1)}}=z_1$.}	
	\label{fig:diagrams}
\end{center}
\end{figure*}

We  consider now the case of a 
random multiplex network with $M=3$  layers. 
The network has Poisson multidegree distributions and averages given by  
\bea
\avg{k^{(1,1,1)}}&=&z_3,\nonumber \\
\avg{k^{(1,1,0)}}&=&\avg{k^{(1,0,1)}}=\avg{k^{(0,1,1)}}=z_2,\nonumber \\
\avg{k^{(1,0,0)}}&=&\avg{k^{(0,1,0)}}=\avg{k^{(0,0,1)}}=z_1.
\eea
In this case, the messages $S^{\vec{m},\vec{n}}(\vec{s}_j)$ only
depend on   
\bea
\mu &=&\sum_{\alpha=1}^M m^{[\alpha]},\nonumber \\
\nu &=&\sum_{\alpha=1}^M n^{[\alpha]},\nonumber \\
\xi &=&\sum_{\alpha=1}^M s_{j\alpha}m^{[\alpha]}.
\eea
Therefore the   fraction of replica nodes in the RMCGC  $S$ is  
determined by the variables
\bea
x_{\mu,\nu}^{(\xi)}=S^{\vec{m},\vec{n}}(\vec{s}_j).
\eea 
The equations that these variables 
 need to satisfy can be described 
 in a  symbolic way by suitable diagrams (see the Supplementary Material for  details in how to read these diagrams).
Diagrams that describe the equations to determine the value of all the  variables
$x_{\mu,\nu}^{(\xi)}$
are presented in  Fig.~\ref{fig:diagrams}.
These  equations read as
\bea
x_{3,3}^{(3)}&=&p^3\left[1-3h_{1,3}+3h_{2,3}-h_{3,3}\right]\nonumber \\ 
x_{3,2}^{(3)}&=&p^2(1-p)\left[1-2h_{1,2}+h_{2,2}\right]+p^3\left[h_{1,3}-2h_{2,3}+h_{3,3}\right]\nonumber \\
x_{2,2}^{(2)}&=&p^2(1-p)\left[1-2h_{1,2}+h_{2,2}\right]+p^3\left[1-2h_{1,3}+h_{2,3}\right]\nonumber\\ 
x_{3,2}^{(2)}&=&x_{2,2}^{(2)}\nonumber \\
x_{1,1}^{(1)}&=&2p^2(1-p)\left[1-2h_{1,2}+h_{2,2}\right]\nonumber \\
&&+p^3\left[1-h_{1,3}-h_{2,3}+h_{3,3}\right]\nonumber \\
x_{2,1}^{(2)}&=&p^2(1-p)\left[h_{1,2}-h_{2,2}\right]+p^2(1-p)\left[1-2h_{1,2}+h_{2,2}\right]\nonumber\\
&&+p^3\left[h_{1,3}-h_{2,3}\right]\nonumber \\
x_{2,1}^{(1)}&=&x_{11}^{(1)}\nonumber\\
x_{3,1}^{(2)}&=&x_{2,1}^{(2)}\nonumber \\
x_{3,1}^{(3)}&=&2p^2(1-p)\left[h_{1,2}-h_{2,2}\right]+p^3\left[h_{2,3}-h_{3,3}\right]\nonumber \\
S&=&x_{1,1}^{(1)}
\label{m_3l}
\eea
where 
\bea
h_{1,3}&=&e^{-z_1x_{1,1}^{(1)}-z_2(2x_{2,2}^{(2)}+2x_{2,1}^{(2)})-z_3(x_{3,3}^{(3)}+2x_{3,2}^{(3)}+x_{3,1}^{(3)})}\nonumber \\
h_{2,3}&=&e^{-2z_1x_{1,1}^{(1)}-z_2(3x_{2,2}^{(2)}+4x_{2,1}^{(2)})-z_3(x_{3,3}^{(3)}+3x_{3,2}^{(3)}+2x_{3,1}^{(3)})}\nonumber \\
h_{3,3}&=&e^{-3z_1x_{1,1}^{(1)}-z_2(3x_{2,2}^{(2)}+6x_{2,1}^{(2)})-z_3(x_{3,3}^{(3)}+3x_{3,2}^{(3)}+3x_{3,1}^{(3)})}\nonumber \\
h_{1,2}&=&e^{-z_1x_{1,1}^{(1)}-z_{2}(x_{2,2}^{(2)}+x_{2,1}^{(2)}+x_{2,1}^{(1)})-z_3(x_{3,2}^{(2)}+x_{3,1}^{(2)})}\nonumber \\
h_{2,2}&=&e^{-2z_1x_{1,1}^{(1)}-z_{2}(x_{2,2}^{(2)}+2x_{2,1}^{(2)}+2x_{2,1}^{(1)})-z_3(x_{3,2}^{(2)}+2x_{3,1}^{(2)})}.
\eea
We note that, in absence of overlap,   i.e., for $z_1=z$,  $z_2=0$ and $z_3=0$,
Eqs.~(\ref{m_3l}) 
reduce Eqs.~(\ref{P1}).

By defining a suitable order of the variables $x_{\mu,\nu}^{(\xi)}$,
it is possible to introduce a 
vector ${\bf x}$ whose elements are the 
variables $x_{\mu,\nu}^{(\xi)}$, and rewrite the Eqs.~(\ref{m_3l}) 
in a matrix form as 
\bea
{\bf G}({\bf x})={\bf 0}.
\label{mg}
\eea
 {The hybrid discontinuous phase transition (characterized by a square root singularity) can be found by imposing
that the system of Eqs.~(\ref{mg})} 
is satisfied together with the condition that the determinant 
of the Jacobian ${\bf J}$ of ${\bf G}({\bf x})$ equals to zero, that is
\bea
{\bf G}({\bf x})={\bf 0} \; \textrm{ and } \;
\det {\bf J}=0.
\eea
Simulation results  of the percolation process for  multiplex networks in this ensemble are presented in Fig.\ref{fig:3layers_average}, 
and they provide clear evidence of a perfect 
agreement to the theoretical prediction.

\begin{figure}[htb]
\begin{center}
	\includegraphics[width=0.7\columnwidth]{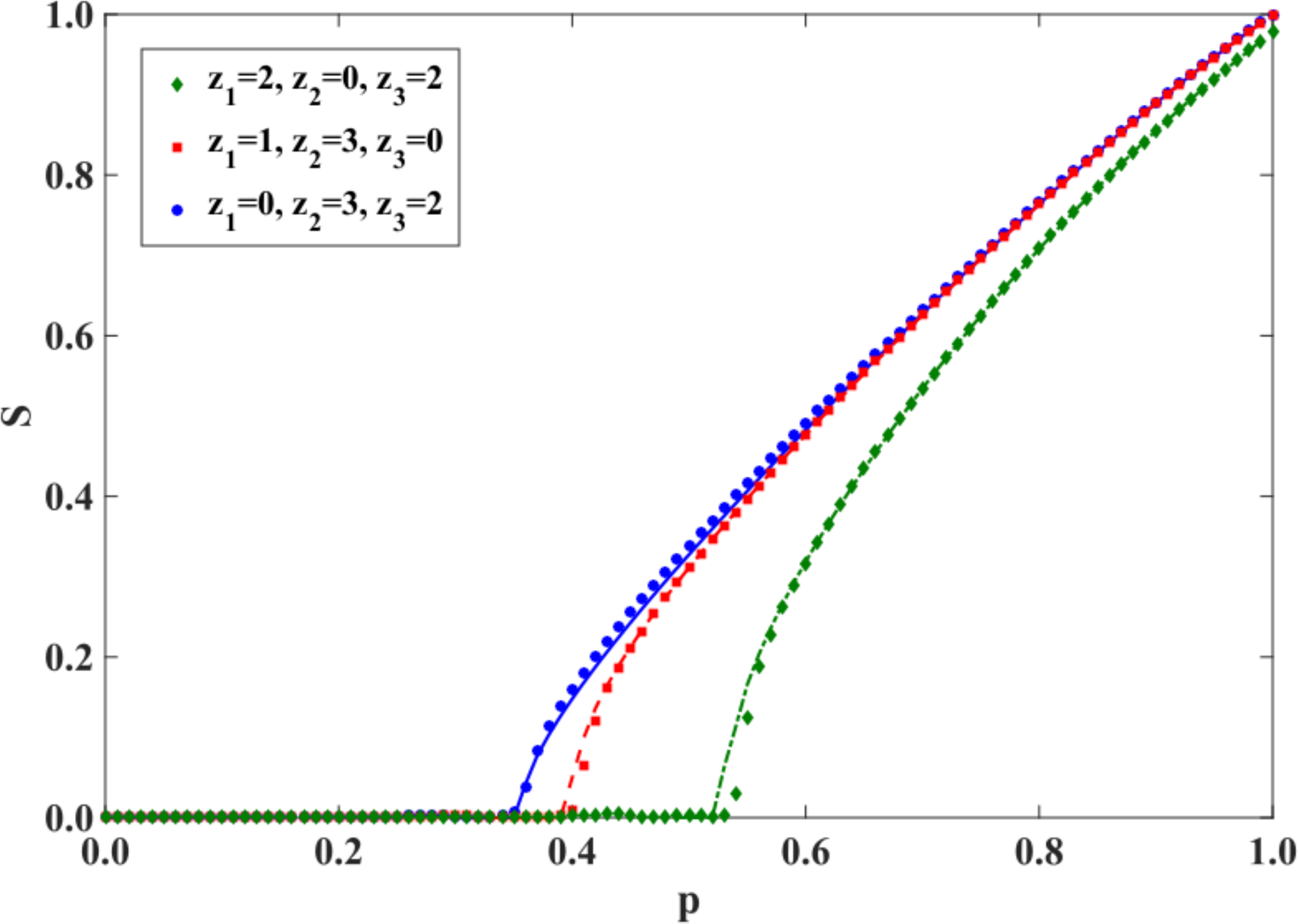}
	\caption{Comparison between the  simulation results of the RMCGC for a multiplex network with $M=3$ layers and Poisson multidegree distribution  with  $\avg{k^{(1,1,1)}}=c_3$, $\avg{k^{(1,1,0)}}=\avg{k^{(1,0,1)}}=\avg{k^{(0,1,1)}}=c_2$ and $\avg{k^{(1,0,0)}}=\avg{k^{(0,1,0)}}=\avg{k^{(0,0,1)}}=c_1$ and the theoretic predictions over the same multilayer network ensemble. Data are shown for $c_1=0,c_2=3,c_3=2$ (blue circles)  $c_1=1,c_2=3,c_3=0$ (red squares) $c_1=2,c_2=0,c_3=2$ (green diamonds). The theoretical predictions are indicated with lines. The simulations results are are performed on multilayer networks with $N=10^4$ nodes and are averaged over $20$ multilayer network realizations.}	
	\label{fig:3layers_average}
\end{center}
\end{figure}

\section{Conclusions}
In this paper, we introduced and fully characterized 
an alternative 
percolation model for multiplex networks. 
The model serves to quantify the robustness of networks with redundant 
interdependencies. According to the model, interdependencies
make a system more fragile than it would be
by considering each layer independently. This fact is 
consistent with the original model used to 
study percolation in multiplex networks~\cite{Havlin1, Son,Baxter2012},
and it is apparent from the
fact that the transition is abrupt for any number 
of network layers
considered
in the interdependent model. On the other hand,
redundancy of interdependencies across multiple layers
favors system robustness, as the height of the discontinuous jump and
the location of the transition point decrease 
as the number of layers increases. This is a fundamental difference
with respect to the model currently adopted to study the robustness of
multiplex networks and multilayer network in general, where instead increasing the number of layers
generates more and more fragile
networks~\cite{Son,Cellai2016,BD1,BD2}.

 {To characterize the model, we deployed a comprehensive
  theoretical
approach based on message passing. Our theory is valid for arbitrary multiplex network
topologies
as long as they are  locally treelike. The theory is
further developed in the context of ensembles of synthetic network models
to analyze properties of the percolation transition emerging in the
new model, and to perform systematic comparisons, and emphasize
fundamental differences, with the
percolation model introduced in Ref.~\cite{Havlin1}. }

 {We remark that ours is not a definitive model, but it
  represents only a good starting point
towards a more realistic description of real  multiplex systems.
For instance, we do not expect to observe a qualitative change of the results if the redundancy is weaker and let us say a node is damaged only if less than two interdependent nodes are functioning. However,
our approach is purely structural and  
in many situations, forgetting about intrinsic dynamical features may
be not appropriate. The primary role of our
model is to emphasize the importance of redundancy or
complementarity in multilayer networks, an obvious-yet-neglected
feature of many real systems.} 
In several realistic settings in fact, system robustness is augmented by
the addition of new layers of interactions, as these
layers are indeed created to provide backup options.
For example, adding a new mode of transportation in a 
preexisting multimodal transportation system should make the
system more reliable against eventual failures. Similarly in a living
organism, the development of new types of interactions among constituents 
should increase the stability of the same organism
against possible mutations. In the current setting, the model
assumes that the functioning of individual nodes requires that 
nodes are correctly operating on at
least two interdependent layers. The model can be, however,
generalized to deal with a variable number of minimal
functioning layers to describe more realistic scenarios in specific 
situations of interest.

\begin{acknowledgements}
F.R. acknowledges support from the National Science Foundation
(CMMI-1552487) and the U.S. Army Research Office (W911NF-16-1-0104).
\end{acknowledgements}



\newpage
\onecolumngrid
\appendix
\newpage

\setcounter{page}{1}
\renewcommand{\theequation}{SM\arabic{equation}}
\setcounter{equation}{0}
\renewcommand{\thefigure}{SM\arabic{figure}}
\setcounter{figure}{0}
\renewcommand{\thetable}{SM\arabic{table}}
\setcounter{table}{0}
\renewcommand{\thesection}{SM\arabic{section}}
\setcounter{section}{0}

\section*{SUPPLEMENTAL MATERIAL}

\subsection{Message passing for given multilayer network and given initial damage}
Let us consider a given multilayer network $\vec{G}$ with $M$ layers. Each layer $\alpha=1,2,\ldots, M$ of the multilayer network has adjacency matrix  ${\bf a}^{[\alpha]}$.
In this multilayer network, each pair of nodes $i$ and $j$ is connected by a multilink 
\bea
\vec{m}_{ij}=(a_{ij}^{[1]},a_{ij}^{[2]}\ldots, a_{ij}^{[\alpha]}, \ldots, a_{ij}^{[M]}).
\eea
Any two nodes $i$ and $j$ are connected by a nontrivial   multilink is  $\vec{m}_{ij}\neq\vec{0}$ implying that at least one link between the two nodes is present across the $M$ layers.
We assume that the initial damage configuration is known and that it is given  by the set of variables $\{s_{i\alpha}\}$ where $s_{i\alpha}$  indicates if a replica $(i,\alpha)$ is initially damaged ($s_{i\alpha}=1$) or not ($s_{i\alpha}=0$).
The message-passing algorithm given in Sec. III of the main text allows us to determine for any given initial damage configuration, if any replica node $(i,\alpha)$ is in the RMCGC ($\sigma_{i\alpha}=1$) or not ($\sigma_{i\alpha}=0$) as long as the multilayer network is locally treelike. 
Specifically the variables  $\sigma_{i\alpha}$ are determined in terms the set of messages 
\bea
\vec{n}_{i\to j}=(n_{i\to j}^{[1]},n_{i\to j}^{[2]},\ldots,n_{i\to j}^{[\alpha]},\ldots n_{i\to j}^{[M]})
\eea
going from  any node $i$ to any node $j$ joined by a nontrivial multilink $\vec{m}_{ij}\neq \vec{0}$.

The messages $\vec{n}_{i\to j}$ are determined according to the following recursive equation 
\bea
n_{i\to j}^{[\alpha]}=\theta(v_{i\to j},2)a^{[\alpha]}_{ij}s_{j\alpha}s_{i\alpha}\left[1-\prod_{\ell\in N_{\alpha}(i)\setminus j}\left(1-n^{[\alpha]}_{\ell\to i}\right)\right],
\label{Rm1}
\eea
where $N_{\alpha}(i)$ indicates the set of nodes that are neighbor of node $i$ in layer $\alpha$ and where $\theta(x)$ is the step function with values $\theta(v_{i \to j},2)=1$ for $v_{i \to j}\geq 2$ and $\theta(v_{i \to j},2)=0$ for $v_{i \to j}=0,1$.
Here the variable  $v_{i \to j}$ indicates in how many layers node $i$ is connected to the RMCGC assuming that node $j$ also belongs to the RMCGC,
\bea
v_{i \to j}&=&\sum_{\alpha=1}^M\left\{s_{i\alpha}\left[1-\prod_{\ell\in N_{\alpha}(i)\setminus j}\left(1-n^{[\alpha]}_{\ell\to i}\right)\right]+s_{i\alpha}s_{j\alpha}a^{[\alpha]}_{ij}\prod_{\ell\in N_{\alpha}(i)\setminus j}\left(1-n^{[\alpha]}_{\ell\to i}\right)\right\}.
\label{Rv}
\eea
Finally the variables $\sigma_{i\alpha}$ are expressed in terms of the messages $\vec{n}_{i\to j}$ and are given by  
\bea
\sigma_{i\alpha}&=&s_{i\alpha}\left[1-\prod_{\ell \in N_{\alpha}(i)}\left(1-n^{[\alpha]}_{\ell\to i}\right)\right]\left\{1-\prod_{\alpha' \neq \alpha}\left[1-s_{i\alpha'}+s_{i\alpha'}\prod_{\ell \in N_{\alpha'}(i)}\left(1-n^{[\alpha']}_{\ell\to i}\right)\right]\right\}.
\label{Rs1}
\eea

In many situations, however, the initial configuration of the damaged $\{s_{i\alpha}\}$ is not known, and instead it is only known the probability distribution $\hat{\mathcal P}( \{s_{i\alpha}\})$ of the initial damage configuration.

In this case, one aims to know the probability $\hat{\sigma}_{i\alpha}=\Avg{\sigma_{i\alpha}}$ that a replica node $(i,\alpha)$ is in the RMCGC for a random configuration of the initial damage.
The value of $\hat{\sigma}_{i\alpha}$, on a locally treelike multilayer network  is determined by a distinct message-passing algorithm that can be derived from the message-passing algorithm valid for single realization of the initial damage, by performing a suitable average of the messages.

Particular care should be  taken when one aims to perform this average.
In fact $\sigma_{i\alpha}$ depends on all the   messages ${n}_{i \to j}^{[\alpha]}$ sent by node $i$ to node $j$ in all the  layers $\alpha$. These messages  are correlated and therefore they cannot be averaged independently.

An alternative  formulation of the Eqs. $(\ref{Rm1})-(\ref{Rps})$ provides the necessary framework for deriving in few steps the message-passing algorithm to predict  $\hat{\sigma}_{i\alpha}$.
This alternative formulation is written terms of  the variables $\sigma_{i\to j}^{\vec{m},\vec{n}}$ indicating whether ($\sigma_{i\to j}^{\vec{m},\vec{n}}=1$) or not ($\sigma_{i\to j}^{\vec{m},\vec{n}}=0$) node $i$ send to node $j$ a message $\vec{n}=\vec{n}_{i\to j}$ given that node $j$ is connected to node $i$ by a multilink $\vec{m}=\vec{m}_{ij}$.

Using Eqs. $(\ref{Rm1})-(\ref{Rv})$ it is easy to see that the  value of the variables $\sigma_{i\to j}^{\vec{m},\vec{n}}$ is determined by the following equations:

\begin{itemize}
\item[(a)]if $\nu=\sum_{\alpha=1}^M n^{[\alpha]}>1$ and $\vec{m}=\vec{m}_{ij}$,
\bea
\hspace{-10mm}\sigma^{\vec{m},\vec{n}}_{i\to j}&=&\prod_{\alpha=1}^M\left[m^{[\alpha]}s_{j\alpha}s_{i\alpha}-m^{[\alpha]}s_{j\alpha}s_{i\alpha}\prod_{\ell\in N(i)\setminus j}\left(1-n^{[\alpha]}_{\ell\to i}\right)\right]^{n^{[\alpha]}}\prod_{\alpha=1}^M\left[1-s_{i\alpha}+s_{i\alpha}\prod_{\ell\in N(i)\setminus j}\left(1-n^{[\alpha]}_{\ell\to i}\right)\right]^{(1-n^{[\alpha]})m^{[\alpha]}s_{j\alpha}},
\label{Ra1}
\eea
\item[(b)]if $\nu=\sum_{\alpha=1}^M n^{[\alpha]}=1$ and $\vec{m}=\vec{m}_{ij}$,
\bea
\hspace{-10mm}\sigma^{\vec{m},\vec{n}}_{i\to j}=\left\{\prod_{\alpha=1}^M\left[m^{[\alpha]}s_{j\alpha}s_{i\alpha}-m^{[\alpha]}s_{j\alpha}s_{i\alpha}\prod_{\ell\in N(i)\setminus j}\left(1-n^{[\alpha]}_{\ell\to i}\right)\right]^{n^{[\alpha]}}\right\}\left\{1-\prod_{\alpha' |n^{[\alpha']}=0}\left[1-s_{i\alpha'}+s_{i\alpha'}\prod_{\ell \in N(i)}\left(1-n^{[\alpha']}_{\ell\to i}\right)\right]\right\},
\label{Ra2}
\eea
\item[(c)]if $\nu=\sum_{\alpha=1}^M n^{[\alpha]}=0$ and $\vec{m}=\vec{m}_{ij}$,
\bea
\sigma^{\vec{m},\vec{0}}_{i\to j}=1-\sum_{\vec{n}\neq \vec{0}}\sigma^{\vec{m},\vec{n}}_{i\to j},
\eea
\end{itemize}
where $\vec{n}_{i\to j}$ is determined in terms of the messages $\sigma^{\vec{m},\vec{n}}_{i\to j}$ as
\bea
\vec{n}_{i\to j}=\mbox{argmax}_{\vec{n}}\sigma^{\vec{m}_{ij},\vec{n}}_{i\to j}.
\eea

Finally a replica node $(i,\alpha)$ is in the RMCGC ($\sigma_{i\alpha}=1$) or not ($\sigma_{i\alpha}=0$) 
depending on the messages it receives from its neighbors, i.e.
\bea
\hspace{-10mm}\sigma_{i\alpha}&=&s_{i\alpha}\left[1-\prod_{\ell \in N(i)}\left(1-n^{[\alpha]}_{\ell\to i}\right)\right]\left\{1-\prod_{\alpha' \neq \alpha}\left[1-s_{i\alpha'}+s_{i\alpha'}\prod_{\ell \in N(i)}\left(1-n^{[\alpha']}_{\ell\to i}\right)\right]\right\}.
\label{Ra3}
\eea

\subsection{Message-passing algorithm  for  random damage}

By averaging Eqs. $(\ref{Ra1})-(\ref{Ra2})-(\ref{Ra3})$ we can derive the message-passing algorithm predicting the probability $\hat{\sigma}_{i\alpha}$ that a replica node $(i,\alpha)$ is in the RMCGC when the initial damage $\{s_i\}$ is randomly drawn for the probability distribution $\hat{\mathcal P}( \{s_{i\alpha}\})$.
Assuming that each replica node is damaged independently the probability distribution $\hat{\mathcal P}( \{s_{i\alpha}\})$ is given by 
\bea
\hat{\mathcal P}( \{s_{i\alpha}\})=\prod_{i=1}^N\prod_{\alpha=1}^M p^{s_{i\alpha}}(1-p)^{1-s_{i\alpha}}.
\label{Rps}
\eea
The  message-passing algorithm valid for a random distribution of the initial disorder, is written in terms of the messages   $\hat{\sigma}_{i\to j}^{\vec{m},\vec{n}}(\vec{s})$. The messages $\hat{\sigma}_{i\to j}^{\vec{m},\vec{n}}(\vec{s})$ take real values between zero and one. They  indicate the probability  that node $i$ send to node $j$ a message $\vec{n}=\vec{n}_{i\to j}$ given that node $j$ is connected to node $i$ by a multilink $\vec{m}=\vec{m}_{ij}$ and that node $j$ has initial damage configuration $\vec{s}=\vec{s}_j$, i.e. $(s_1,s_2,\ldots, s_{\alpha},\ldots, s_M)=(s_{j1},s_{j2},\ldots, s_{jM})$.

Let us indicate with $\hat{P}(\vec{s})$ the probability of a local initial damage configuration given by 
\bea
\hat{P}(\vec{s})=\prod_{\alpha=1}^M p^{s_{\alpha}}(1-p)^{1-s_{\alpha}}
\label{Rlps}
\eea
and let us indicate with $\vec{r}$ the vector 
\bea
\vec{r}=(r^{[1]},r^{[2]},\ldots, r^{[\alpha]}, \ldots, r^{[M]})\eea
of elements $r^{[\alpha]}=0,1$.
Using this notation, the   messages $\hat{\sigma}_{i\to j}^{\vec{m},\vec{n}}(\vec{s})$ are determined by the following algorithm (see last section of this Supplementary Information for the derivation of these results):
\begin{itemize}
\item[(a)]if $\nu=\sum_{\alpha=1}^M n^{[\alpha]}>1$ and $\vec{m}=\vec{m}_{ij}$,
\bea
\hat{\sigma}^{\vec{m},\vec{n}}_{i\to j}(\vec{s})&=& \sum_{\vec{s}_i|\sum_\alpha s_{i\alpha}>1}\hat{P}(\vec{s}_i)\sum_{\vec{r}|r^{[\alpha]}=0 \ \mbox{if}\ \left(n^{[\alpha]}+(1-n^{[\alpha]})m^{[\alpha]}s_{\alpha}\right)=0} {\mathcal C}^{\vec{m},\vec{n}}(\vec{s}_i,\vec{s},\vec{r})\nonumber \\
&&\times\left[\prod_{\ell\in N(i)\setminus j}\left(1-\sum_{\vec{n}'|\sum_{\alpha}(n^{\prime})^{[\alpha]}r^{[\alpha]}>0}\hat{\sigma}_{\ell\to i}^{\vec{m}_{\ell i}\vec{n}'}(\vec{s}_i)\right)\right],
\label{Rb1}
\eea
where
\bea
{\mathcal C}^{\vec{m},\vec{n}}(\vec{s}_i,\vec{s},\vec{r})=\prod_{\alpha=1}^M\left[(m^{[\alpha]}s_{i\alpha}s_{\alpha})^{n^{[\alpha]}}(-1)^{r^{[\alpha]}n^{[\alpha]}}(1-s_{i\alpha})^{(1-r^{[\alpha]})(1-n^{[\alpha]})m^{[\alpha]}s_{\alpha}}\left(s_{i\alpha}\right)^{r^{[\alpha]}(1-n^{[\alpha]})m^{[\alpha]}s_{\alpha}}\right],
\label{RC}
\eea

\item[(b)] if $\nu=\sum_{\alpha'=1}^M n^{[\alpha']}=1$, $n^{[\alpha]}=1$ and $\vec{m}=\vec{m}_{ij}$,
\bea
{\hat{\sigma}}^{\vec{m},\vec{n}}_{i\to j}(\vec{s})&=& \sum_{\vec{s}_i|\sum_{\alpha'} s_{i\alpha'}>1}\hat{P}(\vec{s}_i)s_{i\alpha}s_{\alpha}a_{ij}^{[\alpha]}\left\{
1-\prod_{\ell\in N(i)\setminus j }\left(1-\sum_{\vec{n}'|(n^{\prime})^{[\alpha]}>0}\hat{\sigma}_{\ell\to i}^{\vec{m}_{\ell i}\vec{n}'}(\vec{s}_i)\right)\right.\nonumber\\
&&-\sum_{\vec{r}|r^{[\alpha]}=0}\prod_{\alpha'\neq \alpha}(1-s_{i\alpha'})^{\left(1-r^{\left[\alpha'\right]}\right)}\left(s_{i\alpha'}\right)^{r^{\left[\alpha'\right]}}\prod_{\ell\in N(i)\setminus j}\left(1-\sum_{\vec{n}'|\sum_{\alpha'}(n^{\prime})^{\left[\alpha'\right]}r^{\left[\alpha'\right]}>0}\hat{\sigma}_{\ell\to i}^{\vec{m}_{\ell i}\vec{n}'}(\vec{s}_i)\right)\nonumber \\
&&+\left.\sum_{\vec{r}|r^{[\alpha]}=0}\prod_{\alpha'\neq \alpha}(1-s_{i\alpha'})^{\left(1-r^{\left[\alpha'\right]}\right)}\left(s_{i\alpha'}\right)^{r^{\left[\alpha'\right]}}\prod_{\ell\in N(i)\setminus j}\left(1-\sum_{\vec{n}'|\sum_{\alpha'}(n^{\prime})^{\left[\alpha'\right]}\left[\delta_{\alpha,\alpha'}+r^{\left[\alpha'\right]}\right]>0}\hat{\sigma}_{\ell\to i}^{\vec{m}_{\ell i}\vec{n}'}(\vec{s}_i)\right)\right\},
\label{Rb2}
\eea
\item[(c)]if $\nu=\sum_{\alpha=1}^M n^{[\alpha]}=0$ and $\vec{m}=\vec{m}_{ij}$,
\bea
\hat{\sigma}^{\vec{m},\vec{0}}_{i\to j}(\vec{s})=1-\sum_{\vec{n}\neq \vec{0}}\hat{\sigma}^{\vec{m},\vec{n}}_{i\to j}(\vec{s}).
\eea
\end{itemize}
Finally the probability $\hat{\sigma}_{i\alpha}$ that a replica node $(i,\alpha)$ is in the RMCGC is given by 
\bea
\hat{\sigma}_{i\alpha}&=& \sum_{\vec{s}_i|\sum_{\alpha'} s_{i\alpha'}>1}\hat{P}(\vec{s}_i)s_{i\alpha}a_{ij}^{[\alpha]}\left\{
1-\prod_{\ell\in N(i) }\left(1-\sum_{\vec{n}'|(n^{\prime})^{[\alpha]}>0}\hat{\sigma}_{\ell\to i}^{\vec{m}_{\ell i}\vec{n}'}(\vec{s}_i)\right)\right.\nonumber\\
&&-\sum_{\vec{r}|r^{[\alpha]}=0}\prod_{\alpha'\neq \alpha}(1-s_{i\alpha'})^{\left(1-r^{\left[\alpha'\right]}\right)}\left(s_{i\alpha'}\right)^{r^{\left[\alpha'\right]}}\prod_{\ell\in N(i) }\left(1-\sum_{\vec{n}'|\sum_{\alpha'}(n^{\prime})^{\left[\alpha'\right]}r^{\left[\alpha'\right]}>0}\hat{\sigma}_{\ell\to i}^{\vec{m}_{\ell i}\vec{n}'}(\vec{s}_i)\right)\nonumber \\
&&+\left.\sum_{\vec{r}|r^{[\alpha]}=0}\prod_{\alpha'\neq \alpha}(1-s_{i\alpha'})^{\left(1-r^{\left[\alpha'\right]}\right)}\left(s_{i\alpha'}\right)^{r^{\left[\alpha'\right]}}\prod_{\ell\in N(i)}\left(1-\sum_{\vec{n}'|\sum_{\alpha'}(n^{\prime})^{\left[\alpha'\right]}\left[\delta_{\alpha,\alpha'}+r^{\left[\alpha'\right]}\right]>0}\hat{\sigma}_{\ell\to i}^{\vec{m}_{\ell i}\vec{n}'}(\vec{s}_i)\right)\right\}.
\label{Rb3}
\eea

\subsection{Average over multilayer ensemble with give multidegree sequence}

In order to derive the phase diagram of the percolation transition in presence of redundant interdependencies over given multilayer network ensembles, it is useful to consider a further average of the messages $\hat{\sigma}_{i\to j}^{\vec{m},\vec{n}}$.
To this end we consider the multilayer network ensemble that preserves the multidegree sequence $\{k_i^{\vec{m}}\}$.
Every multilayer network $\vec{G}$ in this ensemble has  probability 
\bea
\tilde{\mathcal P}(\vec{G})=\frac{1}{\tilde{Z}}\prod_{i=1}^N \prod_{\vec{m}\neq \vec{0}}\delta\left[k_i^{\vec{m}},\sum_{j=1}^N \delta\left({\vec{m}},\vec{m}_{ij}\right)\right],
\eea
 where $\tilde{Z}$ is a normalization constant equal to the number of multilayer networks with the given multidegree sequence.

In this multilayer network ensemble the average messages $S^{\vec{m},\vec{n}}(\vec{s})=\Avg{\hat{\sigma}_{i\to j}^{\vec{m}_{ij},\vec{n}}|\vec{m}=\vec{m}_{ij}}$  indicate the probability that a message $\vec{n}$ is sent toward a node with initial damage configuration $\vec{s}$ over a  multilink $\vec{m}$.
These average messages can be found by solving the following recursive equations:
\begin{itemize}
\item[(a)]if $\nu=\sum_{\alpha=1}^M n^{[\alpha]}>1$ 
\bea
S^{\vec{m},\vec{n}}(\vec{s})&=& \sum_{\{k^{\vec{m}}\}}\frac{k^{\vec{m}}}{\Avg{k^{\vec{m}}}}P(\{k^{\vec{m}}\})\sum_{\vec{s}_i|\sum_\alpha s_{i\alpha}>1}\hat{P}(\vec{s}_i)\sum_{\vec{r}|r^{[\alpha]}=0 \ \mbox{if}\ \left(n^{[\alpha]}+(1-n^{[\alpha]})m^{[\alpha]}s_{\alpha}\right)=0} {\mathcal C}^{\vec{m},\vec{n}}(\vec{s}_i,\vec{s},\vec{r})\nonumber \\
&&\times\left[\prod_{\vec{m}'\neq{0}}\left(1-\sum_{\vec{n}'|\sum_{\alpha}(n^{\prime})^{[\alpha]}r^{[\alpha]}>0}S^{\vec{m}'\vec{n}'}(\vec{s}_i)\right)^{k^{\vec{m}'}-\delta(\vec{m},\vec{m}')}\right],
\eea
where
\bea
{\mathcal C}^{\vec{m},\vec{n}}(\vec{s}_i,\vec{s},\vec{r})=\prod_{\alpha=1}^M\left[(m^{[\alpha]}s_{i\alpha}s_{\alpha})^{n^{[\alpha]}}(-1)^{r^{[\alpha]}n^{[\alpha]}}(1-s_{i\alpha})^{(1-r^{[\alpha]})(1-n^{[\alpha]})m^{[\alpha]}s_{\alpha}}\left(s_{i\alpha}\right)^{r^{[\alpha]}(1-n^{[\alpha]})m^{[\alpha]}s_{\alpha}}\right],
\eea

\item[(b)] if $\nu=\sum_{\alpha'=1}^M n^{[\alpha']}=1$, $n^{[\alpha]}=1$ 
\bea
S^{\vec{m},\vec{n}}(\vec{s})&=& \sum_{\{k^{\vec{m}}\}}\frac{k^{\vec{m}}}{\Avg{k^{\vec{m}}}}P(\{k^{\vec{m}}\}) \sum_{\vec{s}_i|\sum_{\alpha'} s_{i\alpha'}>1}\hat{P}(\vec{s}_i)s_{i\alpha}s_{\alpha}a_{ij}^{[\alpha]}\left\{
1-\prod_{\vec{m}'\neq \vec{0} }\left(1-\sum_{\vec{n}'|(n^{\prime})^{[\alpha]}>0}S^{\vec{m}'\vec{n}'}(\vec{s}_i)\right)^{k^{\vec{m}'}-\delta(\vec{m},\vec{m}')}\right.\nonumber\\
&&-\sum_{\vec{r}|r^{[\alpha]}=0}\prod_{\alpha'\neq \alpha}(1-s_{i\alpha'})^{\left(1-r^{\left[\alpha'\right]}\right)}\left(s_{i\alpha'}\right)^{r^{\left[\alpha'\right]}}\prod_{\vec{m}'\neq \vec{0}}\left(1-\sum_{\vec{n}'|\sum_{\alpha'}(n^{\prime})^{\left[\alpha'\right]}r^{\left[\alpha'\right]}>0}S^{\vec{m}'\vec{n}'}(\vec{s}_i)\right)^{k^{\vec{m}}-\delta(\vec{m},\vec{m}')}\nonumber \\
&&\hspace*{-15mm}+\left.\sum_{\vec{r}|r^{[\alpha]}=0}\prod_{\alpha'\neq \alpha}(1-s_{i\alpha'})^{\left(1-r^{\left[\alpha'\right]}\right)}\left(s_{i\alpha'}\right)^{r^{\left[\alpha'\right]}}\prod_{\vec{m}'\neq \vec{0}}\left(1-\sum_{\vec{n}'|\sum_{\alpha'}(n^{\prime})^{\left[\alpha'\right]}\left[\delta_{\alpha,\alpha'}+r^{\left[\alpha'\right]}\right]>0}S^{\vec{m}'\vec{n}'}(\vec{s}_i)\right)^{k^{\vec{m}}-\delta(\vec{m},\vec{m}')}\right\},
\eea
\item[(c)]if $\nu=\sum_{\alpha=1}^M n^{[\alpha]}=0$ 
\bea
S^{\vec{m},\vec{0}}(\vec{s})=1-\sum_{\vec{n}\neq \vec{0}}S^{\vec{m},\vec{n}}(\vec{s}).
\eea
\end{itemize}
Finally the probability $S_{\alpha}$ that a replica node in layer $\alpha$ is in the RMCGC in the multilayer network ensemble is given by 
\bea
S_{\alpha}&=& \sum_{\{k^{\vec{m}}\}} P(\{k^{\vec{m}}\}) \sum_{\vec{s}_i|\sum_{\alpha'} s_{i\alpha'}>1}\hat{P}(\vec{s}_i)s_{i\alpha}a_{ij}^{[\alpha]}\left\{
1-\prod_{\vec{m}'\neq \vec{0} }\left(1-\sum_{\vec{n}'|(n^{\prime})^{[\alpha]}>0}S^{\vec{m}'\vec{n}'}(\vec{s}_i)\right)^{k^{\vec{m}'}}\right.\nonumber\\
&&-\sum_{\vec{r}|r^{[\alpha]}=0}\prod_{\alpha'\neq \alpha}(1-s_{i\alpha'})^{\left(1-r^{\left[\alpha'\right]}\right)}\left(s_{i\alpha'}\right)^{r^{\left[\alpha'\right]}}\prod_{\vec{m}'\neq \vec{0}}\left(1-\sum_{\vec{n}'|\sum_{\alpha'}(n^{\prime})^{\left[\alpha'\right]}r^{\left[\alpha'\right]}>0}S^{\vec{m}'\vec{n}'}(\vec{s}_i)\right)^{k^{\vec{m}}}\nonumber \\
&&+\left.\sum_{\vec{r}|r^{[\alpha]}=0}\prod_{\alpha'\neq \alpha}(1-s_{i\alpha'})^{\left(1-r^{\left[\alpha'\right]}\right)}\left(s_{i\alpha'}\right)^{r^{\left[\alpha'\right]}}\prod_{\vec{m}'\neq \vec{0}}\left(1-\sum_{\vec{n}'|\sum_{\alpha'}(n^{\prime})^{\left[\alpha'\right]}\left[\delta_{\alpha,\alpha'}+r^{\left[\alpha'\right]}\right]>0}S^{\vec{m}'\vec{n}'}(\vec{s}_i)\right)^{k^{\vec{m}}}\right\}.
\eea

\subsection{Derivation of Eq. $(\ref{Rb1})$ }

In this section, we will discuss in detail the derivation of Eq. $(\ref{Rb1})$ from Eq.($\ref{Ra1}$). A similar derivation (that we omit here) can be performed to derive Eqs. $(\ref{Rb2})/(\ref{Rb3})$ from Eqs. $(\ref{Ra2})/(\ref{Ra3})$.

We start from Eq. (\ref{Ra1}) written  for the messages $\sigma^{\vec{m},\vec{n}}_{i\to j}$ sent from a node $i$ to a node $j$ with $\vec{n}$ satisfying 
 $\nu=\sum_{\alpha=1}^M n^{[\alpha]}>1$ and $\vec{m}=(a_{ij}^{[1]},a_{ij}^{[2]}\ldots, a_{ij}^{[M]})$.
 This equation is  rewritten here for convenience,
\bea
\hspace{-10mm}\sigma^{\vec{m},\vec{n}}_{i\to j}&=&\prod_{\alpha=1}^M\left[m^{[\alpha]}s_{j\alpha}s_{i\alpha}-m^{[\alpha]}s_{j\alpha}s_{i\alpha}\prod_{\ell\in N(i)\setminus j}\left(1-n^{[\alpha]}_{\ell\to i}\right)\right]^{n^{[\alpha]}}\prod_{\alpha=1}^M\left[1-s_{i\alpha}+s_{i\alpha}\prod_{\ell\in N(i)\setminus j}\left(1-n^{[\alpha]}_{\ell\to i}\right)\right]^{(1-n^{[\alpha]})m^{[\alpha]}s_{j\alpha}},
\label{Ra1b}
\eea

We  given a set of variables $p^{[\alpha]}=0,1$ we can use the following identity  
\bea
	&&\prod_{\alpha=1}^M (y_{\alpha}+z_{\alpha})^{p^{[\alpha]}}=\prod_{\alpha|p^{[\alpha]}>0}(y_{\alpha}+z_{\alpha}) =\sum_{\vec{r}|r^{[\alpha]}=0 \ \mbox{{\scriptsize if}}\  p^{[\alpha]}=0} \prod_{\alpha=1}^M\left[\left(y_{\alpha}\right)^{1-r^{[\alpha]}} \left(z_{\alpha}\right)^{r^{[\alpha]}}\right],
	\label{r}
\eea
where in the last expression we perform a sum over all the $M$-dimensional vectors $\vec{r}$ 
\bea
\vec{r}=(r^{[1]},r^{[2]},\ldots, r^{[\alpha]}, \ldots,r^{[M]}),
\eea
with $r^{[\alpha]}=0,1$ if $p^{[\alpha]}=1$ and $r^{[\alpha]}=0$ if $p^{[\alpha]}=0$.
Using this expansion for the products in Eq. $(\ref{Ra1b})$ we obtain
\bea
\sigma^{\vec{m},\vec{n}}_{i\to j}&=& \sum_{\vec{r}|r^{[\alpha]}=0 \ \mbox{if}\ \left(n^{[\alpha]}+(1-n^{[\alpha]})m^{[\alpha]}s_{j\alpha}\right)=0}{\mathcal C}^{\vec{m},\vec{n}}(\vec{s}_i,\vec{s},\vec{r})\prod_{\ell\in N(i)\setminus j}\left[\prod_{\alpha=1}^M\left(1-n^{[\alpha]}_{\ell\to i}\right)^{r^{[\alpha]}}\right],
\label{Rd2}
\eea
where ${\mathcal C}^{\vec{m},\vec{n}}(\vec{s}_i,\vec{s},\vec{r})$ is given by Eq. $(\ref{RC})$.
By using the fact that the messages $\sigma^{\vec{m},\vec{n}}_{i\to j}$ take only values zero or one, that that out of all the messages $\sigma^{\vec{m},\vec{n}}_{i\to j}$ from node $i$ to node $j$ only one is actually equal to one, and all the others are zero, we can rewrite Eq. $(\ref{Rd2})$ as
\bea
\hspace*{-7mm}\sigma^{\vec{m},\vec{n}}_{i\to j}= 
\sum_{\vec{r}|r^{[\alpha]}=0 \ \mbox{if}\ \left(n^{[\alpha]}+(1-n^{[\alpha]})m^{[\alpha]}s_{j\alpha}\right)=0}{\mathcal C}^{\vec{m},\vec{n}}(\vec{s}_i,\vec{s},\vec{r})\prod_{\ell\in N(i)\setminus j}\left(1-\sum_{\vec{n}'|\sum_{\alpha}\left(n^{\prime}\right)^{[\alpha]}r^{[\alpha]}>0}\sigma_{\ell\to i}^{\vec{m}_{\ell i}\vec{n}'}\right).\eea
Finally, averaging over the probability distribution $\hat{P}(\vec{s}_i)$ of the configuration $\vec{s}_i$ of the initial damage of node $i$, in the locally treelike approximation  we obtain for the messages $\hat{\sigma}^{\vec{m},\vec{n}}_{i\to j}(\vec{s}_j)$ the Eq. $(\ref{Rb1})$ that we rewrite here for convenience,
\bea
\hat{\sigma}^{\vec{m},\vec{n}}_{i\to j}(\vec{s}_j)&=& \sum_{\vec{s}_i|\sum_\alpha s_{i\alpha}>1}P(\vec{s}_i)\sum_{\vec{r}|r^{[\alpha]}=0 \ \mbox{if}\ \left(n^{[\alpha]}+(1-n^{[\alpha]})m^{[\alpha]}s_{j\alpha}\right)=0}{\mathcal C}^{\vec{m},\vec{n}}(\vec{s}_i,\vec{s},\vec{r})\nonumber \\
&&\prod_{\ell\in N(i)\setminus j}\left(1-\sum_{\vec{n}'|\sum_{\alpha}\left(n^{\prime}\right)^{[\alpha]}r^{[\alpha]}>0}\hat{\sigma}_{\ell\to i}^{\vec{m}_{\ell i}\vec{n}'}(\vec{s}_i)\right).
\eea

 {
\subsection{Ensembles of multilayer networks link overlap and $M=2$ layers}

In the multilayer network case with $M=2$  and link overlap, every replica node is in
the RMCGC if and only if also 
its interdependent node in the other layer is 
in the RMCGC. Therefore, the only 
messages that are different from zero 
are the 
messages $S^{\vec{m},\vec{n}}(\vec{s}_j=(1,1))$ 
sent to nodes $j$  
in state $\vec{s}_j=(1,1)$. Specifically,  
we  consider  the case of a random multilayer network with
Poisson multidegree distributions characterized by the
averages 
\bea
\avg{k^{(1,1)}}&=&z_2,\nonumber \\
\avg{k^{(0,1)}}&=&\avg{k^{(1,0)}}=z_1.
\eea

The messages $S^{\vec{m},\vec{n}}(\vec{s}_j=(1,1))$ only depend on 
the multiplicity of overlap of the multilinks $\vec{m}$ 
and $\vec{n}$ given respectively by  
\bea
\mu &=& \sum_{\alpha=1}^M m^{[\alpha]},\nonumber \\
\nu &=& \sum_{\alpha=1}^M n^{[\alpha]}.
\eea
The  fraction $S$ of replica nodes in the RMCGC is determined by the variables
\bea
x_{\mu,\nu}=S^{\vec{m},\vec{n}}(\vec{s}_j=(1,1)).
\eea 
The value of  $x_{2,2}$ indicates  the probability  that node $i$ to sends a message $\vec{n}=(1,1)$ to its neighbor $j$ with $\vec{s}_j=(1,1)$ connected by a multilink $\vec{m}=(1,1)$.
This fact occurs
if and  only if node $i$ has both replica nodes that are not initially
damaged (which occurs with probability $p^2$) and if at least one
positive message in each layer $\alpha$ is reaching node $i$ from
neighbors different from $j$.
The value of $x_{1,1}$ indicates the probability  that node $i$ 
sends a message $\vec{n}=(1,0)$ to its neighbor $j$ with
$\vec{s}_j=(1,1)$ 
connected by a multilink $\vec{m}=(1,0)$ or equivalently 
sends a message $\vec{n}=(0,1)$ to its neighbor $j$ with $\vec{s}_j=(1,1)$ connected by a multilink $\vec{m}=(0,1)$.
This fact occurs if and only if node $i$ has both replica nodes that
are not initially damaged 
(which occurs with probability $p^2$) and if at least one positive
message
 in each layer $\alpha$ is reaching node $i$ from 
neighboring nodes different from $j$. 
The latter is a necessary condition to have $v_{i \to j}=2$.
The value $x_{2,1}$ indicates the probability  that node $i$  is
sending a message $\vec{n}=(1,0)$ to its 
neighbor $j$ in state  $\vec{s}_j=(1,1)$ and 
connected by a multilink $\vec{m}=(1,1)$ 
or equivalently sends a message $\vec{n}=(0,1)$ to its 
neighbor $j$ in state $\vec{s}_j=(1,1)$ and connected by a multilink $\vec{m}=(1,1)$.
This fact occurs if only if node $i$ has both replica nodes that are
not initially damaged (which occurs with probability $p^2$) and if at
least one positive message is reaching node $i$ in the layer for which
$n^{[\alpha]}=1$ from neighbors different from $j$ and no positive
message
is reaching node $i$ in the layer where $n^{[\alpha]}=1$ 
from neighboring nodes different from node $j$. 
Finally, $S$ is the probability that a replica node  $(i,{\alpha})$ 
is in the RMCGC which implies that
(i) it is not initially damaged, (ii) its replica node in the other
layer is not initially damaged, and (iii) 
at least one positive message reaches  node $i$ in both layers.

The values of the variables $x_{\mu,\nu}$ and $S$ are therefore  
determined by the following set of equations
\begin{equation}
\begin{array}{ll}
x_{2,2}= & p^2\left[1-2e^{-z_1x_{1,1}-z_2(x_{2,2}+x_{2,1})} +e^{-2z_1x_{1,1}-z_2(x_{2,2}+2x_{2,1})}\right]  \\
x_{2,1}= & p^2\left[e^{-z_1x_{1,1}-z_2(x_{2,2} +x_{2,1})} -e^{-2z_1x_{1,1}-z_2(x_{2,2}+2x_{2,1})}\right] \\
S= & x_{1,1}=x_{2,2}
\end{array} \;.
\end{equation}

These equations are the same equations as those that determine the 
value of the MCGC as long 
as the fact that the damage in each replica node is independent 
is taken into account, which can be done by making the substitution $p^2\to p$~\cite{Baxter2016,Cellai2016}.

\subsection{Ensembles of multilayer networks link overlap and $M=3$ layers}
Let us consider  the case of a 
random multilayer network with $M=3$  layers with Poisson multidegree distributions and averages given by  
\bea
\avg{k^{(1,1,1)}}&=&z_3,\nonumber \\
\avg{k^{(1,1,0)}}&=&\avg{k^{(1,0,1)}}=\avg{k^{(0,1,1)}}=z_2,\nonumber \\
\avg{k^{(1,0,0)}}&=&\avg{k^{(0,1,0)}}=\avg{k^{(0,0,1)}}=z_1.
\eea
In this case, the messages $S^{\vec{m},\vec{n}}(\vec{s}_j)$ only
depend on the
 multiplicity of overlap of the multilinks $\vec{m}$ 
and $\vec{n}$ and the number of layers where $s_{j,\alpha}=1$ and
$m^{[\alpha]}=1$. 
Therefore, messages depend only on  
\bea
\mu &=&\sum_{\alpha=1}^M m^{[\alpha]},\nonumber \\
\nu &=&\sum_{\alpha=1}^M n^{[\alpha]},\nonumber \\
\xi &=&\sum_{\alpha=1}^M s_{j\alpha}m^{[\alpha]}.
\eea
\begin{figure}[htb]
\begin{center}
	\includegraphics[width=0.8\columnwidth]{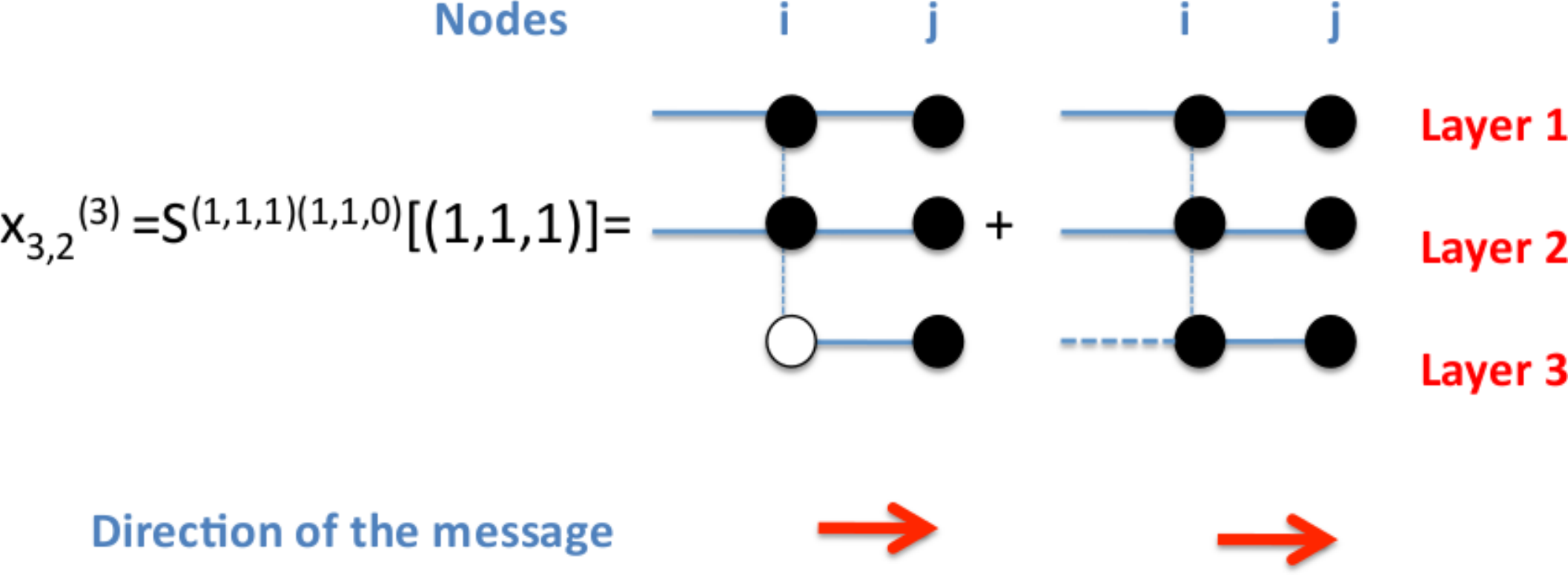}	
	\caption{Example of a diagrammatic representation of the equations determining $x_{3,2}^{(3)}=S^{(1,1,1),(1,1,0)}[(1,1,1)]$ in a multilayer network with $M=3$ layers and $\avg{k^{(1,1,1)}}=z_3$, $\avg{k^{(1,1,0)}}=\avg{k^{(1,0,1)}}=\avg{k^{(0,1,1)}}=z_2$ and $\avg{k^{(1,0,0)}}=\avg{k^{(0,1,0)}}=\avg{k^{(0,0,1)}}=z_1$. 
Filled circles indicate initially undamaged replica nodes $s_{i\alpha}=1$, whereas empty circles indicate initially damaged replica replica nodes $s_{i\alpha}=0$. The message are sent along the direction indicated by the arrows. A solid line reaching node $i$ in layer $\alpha$ indicates that at least one positive message is reaching node $i$ from nodes different from node $j$ in  layer $\alpha$. Dotted lines joining node $i$ in layer $\alpha$ indicate that no positive message reaches node $i$ from nodes different from node $j$ in  layer $\alpha$. A solid (dotted) line between node $i$ and node $j$ in layer $\alpha$ indicates $m^{[\alpha]}=1$ (${m}^{[\alpha]}=0$).}
	\label{Sfig:diagram_example}
\end{center}
\end{figure}
The   fraction of replica nodes in the RMCGC  $S$ is  
determined by the variables
\bea
x_{\mu,\nu}^{(\xi)}=S^{\vec{m},\vec{n}}(\vec{s}_j).
\eea 
\begin{figure*}[htb]
\begin{center}
	\includegraphics[width=0.8\columnwidth]{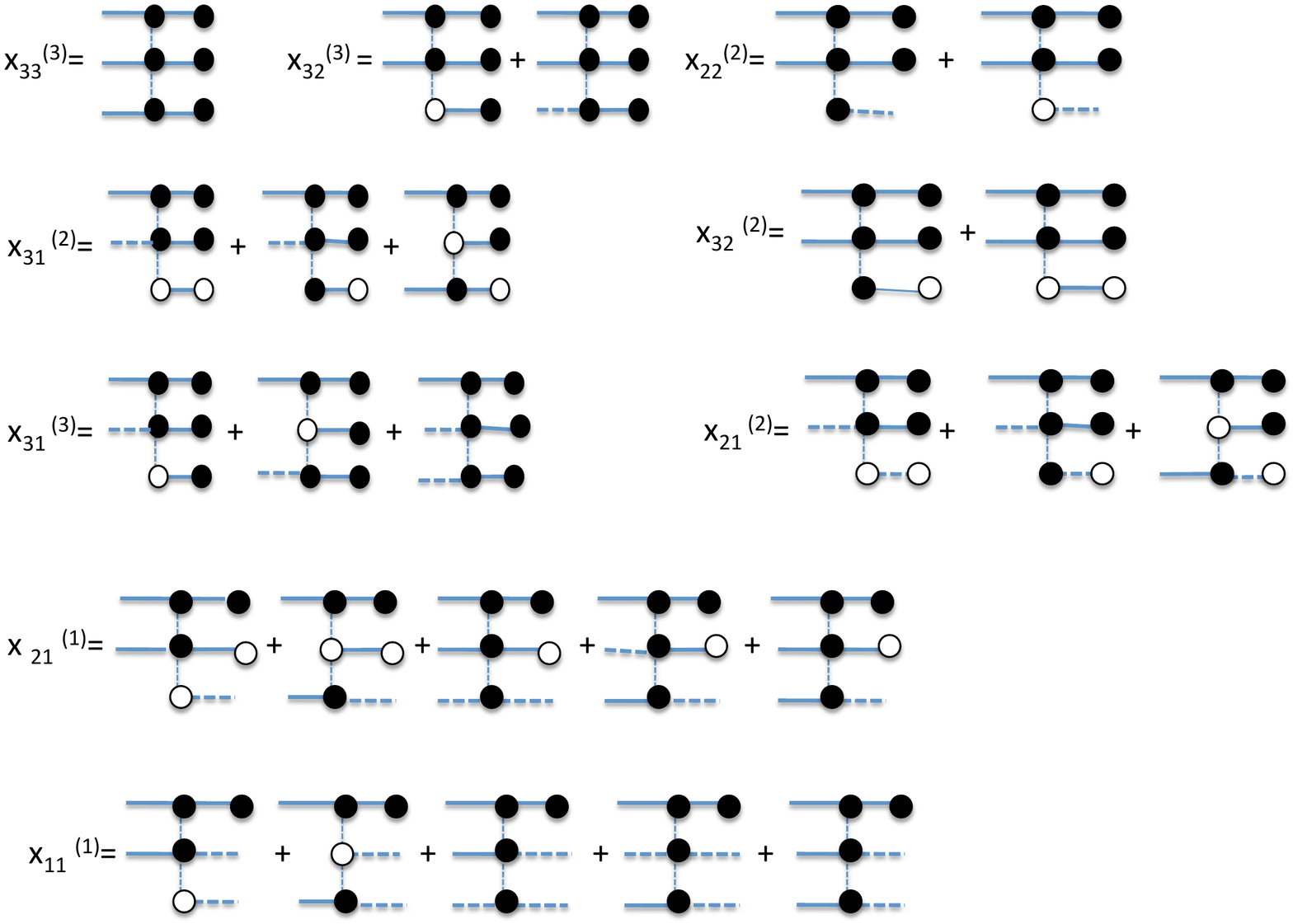}
	\caption{Diagrams for Eqs. $(\ref{m_3l})$ determining $x_{\mu
            \nu}^{(\xi)}$ in the case of multilayer networks with three layers ($M=3$) and Poisson multidegree distribution with $\avg{k^{(1,1,1)}}=z_3$, $\avg{k^{(1,1,0)}}=\avg{k^{(1,0,1)}}=\avg{k^{(0,1,1)}}=z_2$ and $\avg{k^{(1,0,0)}}=\avg{k^{(0,1,0)}}=\avg{k^{(0,0,1)}}=z_1$.}	
	\label{Sfig:diagrams}
\end{center}
\end{figure*}
Let us explicitly describe the equations that one of these variables 
 needs to satisfy, and introduce a symbolic way to describe the
 equations. 
Specifically, we consider $x_{3,2}^{(3)}$ as the probability
$S^{(1,1,1),(1,1,0)}((1,1,1))$ that a node $i$, connected to  a node
$j$ by a multilink $\vec{m}=(1,1,1)$, sends to node $j$ a message
$\vec{n}=(1,1,0)$ 
provided  that  node $j$ is in the state $\vec{s}_j=(1,1,1)$ (see Fig.~\ref{Sfig:diagram_example}).
This probability is equal to the sum of (i)  the probability that  
node $i$ is in the state $\vec{s}_i=(1,1,0)$ [which occurs with
probability $(1-p)p^2$] and it sends the message $\vec{n}=(1,1,0)$  to
node $j$  and (ii) the probability that node $i$ is in the state 
$\vec{s}_i=(1,1,1)$ (which occur with probability $p^3$) and sends the same message to node $j$.
Node $i$ sends the message  $\vec{n}=(1,1,0)$ only if the following conditions are met:
\begin{itemize}
\item[(i)] if node $i$ is in the state $\vec{s}_i=(1,1,0)$,  node $i$ must receive at least one positive message from nodes different from node $j$ in layers $\alpha=1$ and $\alpha=2$.
\item[(ii)]
if node $i$ is in the state $\vec{s}_i=(1,1,1)$, node $i$ must receive at least one positive message from nodes different from node $j$ in layers $\alpha=1$ and $\alpha=2$
and must  not receive any positive message from nodes different from node $j$ in layer $\alpha=3$.
\end{itemize} 
These requirements are summarized by the diagram of Fig.~\ref{Sfig:diagram_example}.
Diagrams that describe the equations to determine the value of all the other variables
$x_{\mu,\nu}^{(\xi)}$
are presented in  Fig.~\ref{Sfig:diagrams}.
These  equations read as
\bea
x_{3,3}^{(3)}&=&p^3\left[1-3h_{1,3}+3h_{2,3}-h_{3,3}\right]\nonumber \\ 
x_{3,2}^{(3)}&=&p^2(1-p)\left[1-2h_{1,2}+h_{2,2}\right]+p^3\left[h_{1,3}-2h_{2,3}+h_{3,3}\right]\nonumber \\
x_{2,2}^{(2)}&=&p^2(1-p)\left[1-2h_{1,2}+h_{2,2}\right]+p^3\left[1-2h_{1,3}+h_{2,3}\right]\nonumber\\ 
x_{3,2}^{(2)}&=&x_{2,2}^{(2)}\nonumber \\
x_{1,1}^{(1)}&=&2p^2(1-p)\left[1-2h_{1,2}+h_{2,2}\right]+p^3\left[1-h_{1,3}-h_{2,3}+h_{3,3}\right]\nonumber \\
x_{2,1}^{(2)}&=&p^2(1-p)\left[h_{1,2}-h_{2,2}\right]+p^2(1-p)\left[1-2h_{1,2}+h_{2,2}\right]+p^3\left[h_{1,3}-h_{2,3}\right]\nonumber \\
x_{2,1}^{(1)}&=&x_{11}^{(1)}\nonumber\\
x_{3,1}^{(2)}&=&x_{2,1}^{(2)}\nonumber \\
x_{3,1}^{(3)}&=&2p^2(1-p)\left[h_{1,2}-h_{2,2}\right]+p^3\left[h_{2,3}-h_{3,3}\right]\nonumber \\
S&=&x_{1,1}^{(1)}
\label{m_3l}
\eea
where 
\bea
h_{1,3}&=&e^{-z_1x_{1,1}^{(1)}-z_2(2x_{2,2}^{(2)}+2x_{2,1}^{(2)})-z_3(x_{3,3}^{(3)}+2x_{3,2}^{(3)}+x_{3,1}^{(3)})}\nonumber \\
h_{2,3}&=&e^{-2z_1x_{1,1}^{(1)}-z_2(3x_{2,2}^{(2)}+4x_{2,1}^{(2)})-z_3(x_{3,3}^{(3)}+3x_{3,2}^{(3)}+2x_{3,1}^{(3)})}\nonumber \\
h_{3,3}&=&e^{-3z_1x_{1,1}^{(1)}-z_2(3x_{2,2}^{(2)}+6x_{2,1}^{(2)})-z_3(x_{3,3}^{(3)}+3x_{3,2}^{(3)}+3x_{3,1}^{(3)})}\nonumber \\
h_{1,2}&=&e^{-z_1x_{1,1}^{(1)}-z_{2}(x_{2,2}^{(2)}+x_{2,1}^{(2)}+x_{2,1}^{(1)})-z_3(x_{3,2}^{(2)}+x_{3,1}^{(2)})}\nonumber \\
h_{2,2}&=&e^{-2z_1x_{1,1}^{(1)}-z_{2}(x_{2,2}^{(2)}+2x_{2,1}^{(2)}+2x_{2,1}^{(1)})-z_3(x_{3,2}^{(2)}+2x_{3,1}^{(2)})}.
\eea
In the main text we compared the result of these equation with simulation results of the percolation process in this ensemble of multiplex networks averaged over different multiplex network realizations. Here for completeness we also show simulation results 
of the percolation process on a single instance of multiplex network and given initial damage in this ensemble with the results of the message-passing algorithm  proposed in Sec. III of the main text (see figure $\ref{fig:MP_3layers}$).

\begin{figure}[htb]
\begin{center}
	\includegraphics[width=0.7\columnwidth]{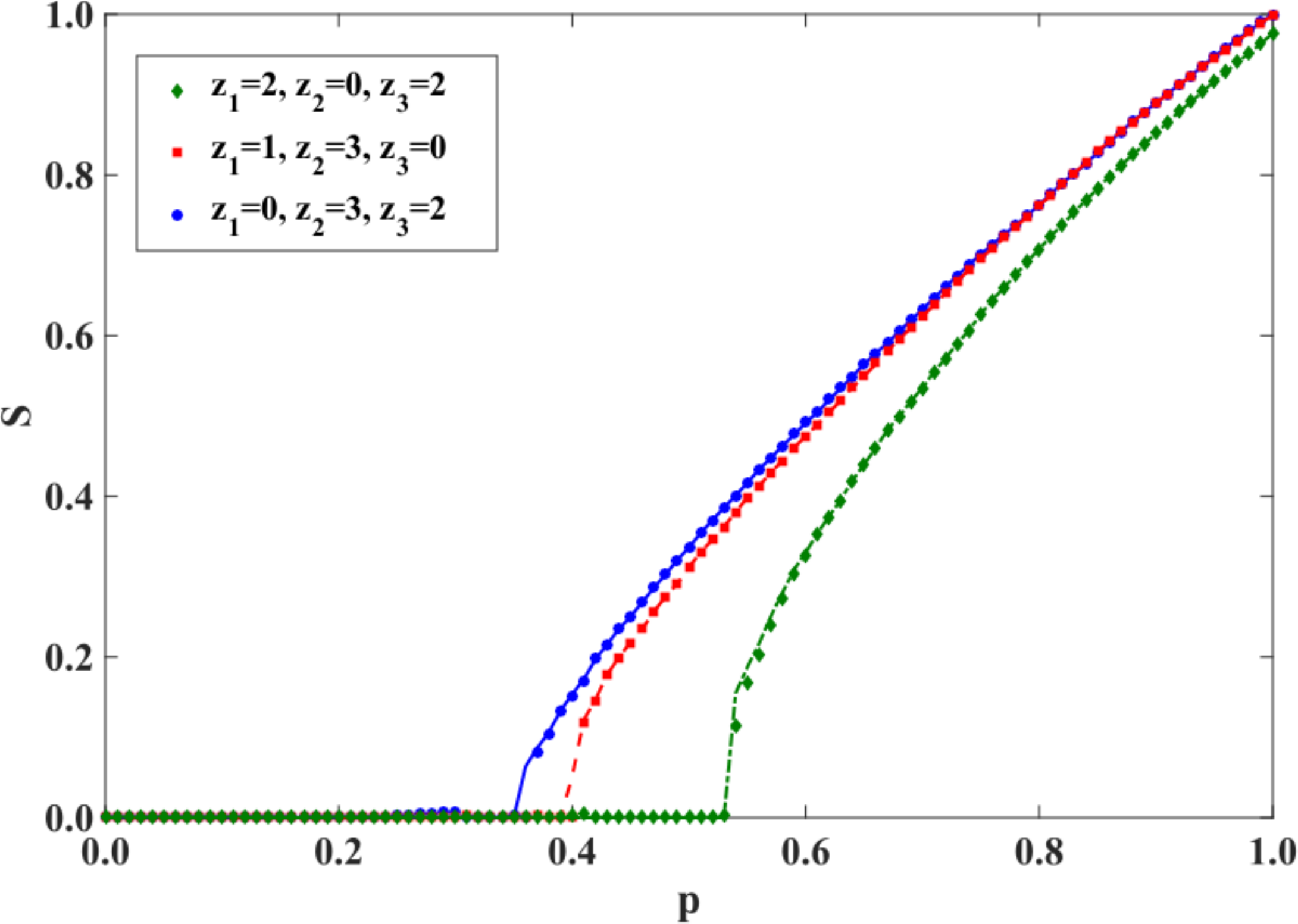}
	\caption{Comparison between the  simulation results 
and message-passing theory for a multiplex network with $M=3$ layers
and Poisson multidegree distribution  with  $\avg{k^{(1,1,1)}}=z_3$,
$\avg{k^{(1,1,0)}}=\avg{k^{(1,0,1)}}=\avg{k^{(0,1,1)}}=z_2$ and
$\avg{k^{(1,0,0)}}=\avg{k^{(0,1,0)}}=\avg{k^{(0,0,1)}}=z_1$. 
We consider here a single network instance and a given 
configuration of damage. Data are shown for $z_1=0,z_2=3,z_3=2$
(blue),  $z_1=1,z_2=3,z_3=0$, (red), and $z_1=2,z_2=0,z_3=2$
(green). Symbols stand for results from numerical simulations,
whereas lines represent results for the numerical solution
of the message-passing equations. Simulations results are performed on
networks with $N=10^4$ nodes.}	
	\label{fig:MP_3layers}
\end{center}
\end{figure}
}


\end{document}